
\global\newcount\meqno
\def\eqn#1#2{\xdef#1{(\secsym\the\meqno)}
\global\advance\meqno by1$$#2\eqno#1$$}
%
\global\newcount\refno
\def\ref#1{\xdef#1{[\the\refno]}
\global\advance\refno by1#1}
\global\refno = 1
\vsize=7.5in
\hsize=5.6in
\magnification=1200
\tolerance 10000
\baselineskip=0.1cm
\baselineskip 12pt plus 1pt minus 1pt
\vskip 2in
\medskip
\centerline{\bf OPERATOR CUTOFF REGULARIZATION AND RENORMALIZATION GROUP}
\bigskip
\medskip
\centerline{\bf IN YANG-MILLS THEORY}
\vskip 24pt
\centerline{Sen-Ben Liao\footnote{$^\dagger$}{e-mail:senben@phy.duke.edu}}
\vskip 12pt
\centerline{\it Department of Physics}
\centerline{\it Duke University }
\centerline{\it Durham, North Carolina\ \ 27708\ \ \ U.S.A.}
\vskip 12pt
\vskip 24 pt
\vskip 2in
\baselineskip 12pt plus 2pt minus 2pt
\centerline{{\bf ABSTRACT}}
\medskip
\medskip

We derive a manifestly gauge invariant low energy blocked action
for Yang-Mills theory using operator cutoff regularization, a prescription
which renders the theory finite with a regulating smearing function
constructed for the proper-time integration. By embedding the momentum cutoff
scales in the smearing function, operator cutoff formalism allows for a
direct application of Wilson-Kadanoff renormalization group to
Yang-Mills theory in a completely gauge symmetry preserving manner.
In particular, we obtain a renormalization group flow equation which
takes into consideration the contributions of higher
dimensional operators and provides a systematic way of exploring the influence
of these operators as the strong coupling, infrared limit is approached.

\vskip 12pt
\vfill
\noindent DUKE-TH-94-65\hfill November 1995
\eject

\vskip 2in
\centerline{\bf I. INTRODUCTION}
\medskip
\nobreak
\xdef\secsym{1.}\global\meqno = 1
\medskip

An important technical issue in quantum
field theory is regularization, the removal of divergences
which arise from incorporating the effects of quantum fluctuations.
Although various regularization schemes are available to make
the theory finite and well defined, it is crucial to subtract off
the infinities with a procedure that preserves all the symmetries of
the original theory. For example, when the underlying
theory possesses gauge symmetry,
prescriptions such as dimensional regularization \ref\dimreg, $\zeta$
function regularization \ref\dowker, invariant Pauli-Villars procedure
\ref\pauli\ and the proper-time regularization \ref\schwinger\ are the
ideal candidates since they respect gauge symmetry; a sharp momentum cut-off,
on the other hand, is not suitable owing to its noninvariant nature.
However, for systems having symmetry properties that are dimensionality
dependent (e.g. chiral symmetry or supersymmetry), it becomes problematic
to use dimensional regularization.

Operator cutoff regularization was proposed in \ref\michael\
as an invariant prescription which simulates the feature
of momentum cutoff regulator. In this formalism, the one-loop
contribution to the effective action is written as \michael\ \ref\ocr\
\eqn\hktw{ {\rm Tr}_{\rm oc}\Bigl({\rm ln}{\cal H}-{\rm ln}{\cal H}_0\Bigr)
= -\int_0^\infty{ds\over s}\rho_k^{(d)}(\Lambda,s)~{\rm Tr}\Bigl(e^{-{\cal H}s}
-e^{-{\cal H}_0s}\Bigr) ,}
where ${\cal H}$ is an arbitrary fluctuation operator governing the
quadratic fluctuations of the fields and ${\cal H}_0$
its corresponding limit of zero background field. The subscript ``oc''
implies that the trace sum is to be operator cutoff regularized using the $d$
dimensional smearing function
${\rho_k^{(d)}(s,\Lambda)}$ which contains both the ultraviolet(UV) cutoff
$\Lambda$ and
the infrared (IR) cutoff $k$. We require ${\rho_k^{(d)}(s,\Lambda)}$ to satisfy
the following
conditions: (1) $\rho_k^{(d)}(s=0,\Lambda)=0$, i.e.,
it must vanish sufficiently fast near $s=0$ to eliminate
the unwanted UV divergence; (2) $\rho_{k=0}^{(d)}(s\to\infty,\Lambda)=1$
since the physics in the IR ($s\sim \infty$) regime is to remain unmodified;
and (3) $\rho^{(d)}_{k=\Lambda}(s,\Lambda)=0$ so that the one-loop correction
to the effective action vanishes at the UV cutoff.
In addition, we have
\eqn\smew{ \rho_{k=0}^{(d)}(s,\Lambda\to\infty)=1,}
which reduces the operator cutoff regularization to the original
Schwinger's proper-time formalism \schwinger:
\eqn\hkor{{\rm Tr}\Bigl({\rm ln}{\cal H}-{\rm ln}{\cal H}_0\Bigr)
=-\int_0^\infty{ds\over s}~{\rm Tr}\Bigl(e^{-{\cal H}s}
-e^{-{\cal H}_0s}\Bigr).}
Thus, operator cutoff may be regarded as a special case of
proper-time regularization. In fact, various
other prescriptions such as sharp proper-time
cut-off, point-splitting method, Pauli-Villars regulator, operator
cutoff, dimensional regularization
and $\zeta$ function regularization all belong to
the generalized class of
proper-time and can be represented by a suitably chosen
smearing function \ref\chiral\ \ref\zuk.

What are the advantages of employing operator cutoff regularization?
A remarkable feature of this regulator is that it contains momentum
cutoff scales yet preserves gauge symmetry. This makes it possible
to examine the renormalization group (RG) flow of gauge theories
using the Wilson-Kadanoff approach \ref\wilson. In the momentum
space formulation of RG, blocking transformation is applied to give the theory
an IR cutoff
scale $k$ that separates the fast-fluctuating modes from
the slowly-varying components \ref\blo; successive elimination of the fast
modes then leads to a low-energy effective blocked action from which the
RG flow pattern can be obtained. The noninvariant nature of
this momentum RG approach is precisely compensated by
invoking operator cutoff formalism
where the procedure of blocking is readily taken over
by the smearing function ${\rho_k^{(d)}(s,\Lambda)}$.

As demonstrated in \michael\ and \ocr, when choosing the
smearing function to be of the form
\eqn\rro{\eqalign{ {\rho_k^{(d)}(s,\Lambda)}
&=\rho^{(d)}(\Lambda^2s)-\rho^{(d)}(k^2s)
={2s^{d/2}\over\Gamma(d/2)}\int_k^{\Lambda}dz~ z^{d-1}
e^{-z^2s}={2s^{d/2}\over{S_d\Gamma(d/2)}}\int_z^{'}e^{-z^2s} \cr
&
={1\over\Gamma(d/2)}\Gamma[{d\over 2};k^2s,\Lambda^2s],}}
with
\eqn\ssd{\int_z=\int{d^dz\over{(2\pi)^d}},\qquad\int_z^{'}=S_d
\int_k^{\Lambda}dz~z^{d-1},
\qquad S_d={2\over{(4\pi)^{d/2}\Gamma(d/2)}},}
the characteristic of sharp momentum cutoff is reproduced in the leading
order blocked potential in derivative expansion.
For the higher order (covariant) derivative terms,
${\rho_k^{(d)}(s,\Lambda)}$ corresponds to a smooth regulator.

In the present paper, we follow the preliminary work of \ocr\ and \ref\ym\
and further extend the use of operator
cutoff to the non-abelian Yang-Mills theory.
The gauge-invariant effective blocked action for the theory is
calculated in covariant background formalism.
Since the physics in the low energy regime is
dominated by nonperturbative effects,
nonperturbative methods such as lattice calculation must be employed.
Our RG equation based on the Wilson-Kadanoff blocking transformation
offers a powerful alternative for probing the physical phenomena
near the confining scale. In the past, various attempts were made
to construct
the Wilson-Kadanoff RG formalism for gauge theories \ref\wkrg\ - \ref\rt.
However, the constraint of gauge symmetry has been the major obstacle
for the success. In \wkrg, the Slavnov-Taylor identities are
perturbatively restored by appropriate choice of boundary conditions for
the RG flow equation. On the other hand, in the average action approach \rt,
additional cutoff-dependent interactions are introduced
in such a way that the overall
average action maintains its gauge invariant form. The spirit of our blocking
transformation formalism is parallel to the latter, however, it is
implemented in the operator cutoff formalism by a smearing function. The
advantage of this approach is that gauge symmetry can be seen to persist
in a rather transparent manner.

The organization of the paper is as follows: In Sec. II, we review
some essential features of the operator cutoff regularization
and illustrate how it is used in conjunction with covariant derivative
expansion. Similarity can be found between our formalism and
the method of higher (covariant) derivatives \ref\faddeev.
Details of computing the gauge-invariant Yang-Mills
blocked action perturbatively in the covariant background formalism are
given in Sec. III. The RG pattern of the blocked action and
the $\beta$ function which
governs the evolution of the coupling constant are investigated in Sec IV.
In Sec. V we apply operator cutoff formalism to examine the
the flow of the theory in a constant chromomagnetic
field configuration. Since the theory develops an imaginary part which signals
instability of the vacuum as the momentum scale
falls below $\sqrt{gB}$, we choose the IR cutoff to be such that
$k^2>gB$, thereby eliminating the difficulties associated with an unstable
vacuum.
An improved RG equation is proposed to take into account
the higher dimensional operators such as ${(B^2/2)}^2$. Section VI is
reserved for summary and
discussions. In Appendix A we provide the details of calculating
the blocked potentials for $\lambda\phi^4$ theory and
scalar electrodynamics in $d=4$ using operator cutoff regularization.
In Appendix B, we compare and contrast
various prescriptions that belong to the
generalized class of proper-time regularization. In particular, we show
how dimensional regularization can
be modified to incorporate cutoff scales. Connection between momentum
regulator and dimensional regularization is readily established in our
``dimensional cutoff'' scheme. Momentum cutoff scales can also be
brought into the $\zeta$ function regularization in a similar manner.

\bigskip
\goodbreak
\medskip
\centerline{\bf II. OPERATOR CUTOFF REGULARIZATION}
\medskip
\nobreak
\xdef\secsym{2.}\global\meqno = 1
\medskip

As mentioned in the Introduction, operator cutoff regularization not only
allows us to bypass the complications of dealing with
divergences, it also encompasses the features of momentum space blocking
transformation in a symmetry-preserving manner.
We first review some essential properties of the
formalism already developed in \ocr.

With the smearing function ${\rho_k^{(d)}(s,\Lambda)}$ written in \rro, the
operator cutoff regularized propagator and the one-loop contribution of
the blocked action become, respectively,
\eqn\hkk{\eqalign{ {1\over {\cal H}^n}\Big\vert_{\rm oc}&=
{1\over\Gamma(n)}
\int_0^\infty ds~s^{n-1}\rho_k^{(d)}(s,\Lambda)e^{-{\cal H}s} \cr
&
={1\over{\cal H}^n}\cdot{{2\Gamma(n+d/2)}\over {d\Gamma(n)\Gamma(d/2)}}
\Biggl\{\Bigl({\Lambda^2\over{\cal H}}\Bigr)^{d/2}F\Bigl({d\over 2},
{d\over 2}+n,1+{d\over 2};-{\Lambda^2\over{\cal H}}\Bigr) \cr
&\qquad\qquad\qquad\qquad~~
-\Bigl({k^2\over{\cal H}}\Bigr)^{d/2}F\Bigl({d\over 2},
{d\over 2}+n,1+{d\over 2};-{k^2\over{\cal H}}\Bigr)\Biggr\},}}
and
\eqn\hkorr{\eqalign{{\rm Tr}_{\rm oc}\Bigl({\rm ln}{\cal H}
-&{\rm ln}{\cal H}_0\Bigr)
=-\int_0^\infty{ds\over s}{\rho_k^{(d)}(s,\Lambda)}{\rm Tr}\Bigl(e^{-{\cal H}s}
-e^{-{\cal H}_0s}\Bigr) \cr
&
=-{2\over d}{\rm Tr}\Biggl\{\Bigl({\Lambda^2\over{\cal H}}\Bigr)^{d/2}F\Bigl(
{d\over 2},
{d\over 2},1+{d\over 2};-{\Lambda^2\over{\cal H}}\Bigr)
-\Bigl({\Lambda^2\over{\cal H}_0}\Bigr)^{d/2}F\Bigl({d\over 2},
{d\over 2},1+{d\over 2};-{\Lambda^2\over{\cal H}_0}\Bigr) \cr
&\qquad~~~
-\Bigl({k^2\over{\cal H}}\Bigr)^{d/2}F\Bigl({d\over 2},
{d\over 2},1+{d\over 2};-{k^2\over{\cal H}}\Bigr)
+\Bigl({k^2\over{\cal H}_0}\Bigr)^{d/2}F\Bigl({d\over 2},
{d\over 2},1+{d\over 2};-{k^2\over{\cal H}_0}\Bigr)\Biggr\},}}
where
\eqn\funn{ F\Bigl(a,b,c;\beta\Bigr)=B^{-1}(b,c-b)\int_0^1 dx~x^{b-1}
(1-x)^{c-b-1}(1-\beta x)^{-a}}
is the hypergeometric function symmetric under the exchange between $a$
and $b$, and
\eqn\ber{ B(x,y)={\Gamma(x)\Gamma(y)\over\Gamma(x+y)}=\int_0^1 dt~t^{x-1}
(1-t)^{y-1},}
the Euler $\beta$ function. For $n=1$ and $2$, eq.\hkk\ gives, respectively,
\eqn\hkko{ {1\over {\cal H}}\Big\vert_{\rm oc}=
\int_0^\infty ds~\rho_k^{(d)}(s,\Lambda)e^{-{\cal H}s}
={1\over{\cal H}}\Biggl\{\Bigl({\Lambda^2\over
{{\cal H}+\Lambda^2}}\Bigr)^{d/2}-\Bigl({k^2\over{{\cal H}+k^2}}\Bigr)^{d/2}
\Biggr\},}
and
\eqn\hkkoy{\eqalign{ {1\over {\cal H}^2}\Big\vert_{\rm oc}&=
\int_0^\infty ds~s\rho_k^{(d)}(s,\Lambda)e^{-{\cal H}s} \cr
&
={1\over{\cal H}^2}\Biggl\{\Bigl({\Lambda^2\over
{{\cal H}+\Lambda^2}}\Bigr)^{d/2}\Bigl[1+{d\over 2}{{\cal H}\over{{\cal
H}+\Lambda^2}}
\Bigr]-\Bigl({k^2\over{{\cal H}+k^2}}\Bigr)^{d/2}\Bigl[1+{d\over 2}{{\cal
H}\over
{{\cal H}+k^2}}\Bigr]\Biggr\}.}}
Consider $d=4$ where
\eqn\rty{ \rho_k^{(4)}(s,\Lambda)=(1+k^2s)e^{-k^2s}-(1+\Lambda^2s)
e^{-\Lambda^2s}.}
We have
\eqn\cher{{1\over {\cal H}}\Big\vert_{\rm oc}={1\over{{\cal
H}+k^2}}-{1\over{{\cal H}
+\Lambda^2}}-{\Lambda^2\over{({\cal H}+\Lambda^2)^2}}+{k^2\over {({\cal
H}+k^2)^2}},}
and
\eqn\hko{\eqalign{ {\rm Tr}_{\rm oc}\Bigl({\rm ln}{\cal H}
-{\rm ln}{\cal H}_0\Bigr)&= {\rm Tr}\Biggl\{{\rm ln}\Bigl[{{{\cal H}+k^2}\over
{{\cal H}_0+k^2}}\times
{{{\cal H}_0+\Lambda^2}\over {{\cal H}+\Lambda^2}}\Bigr]
-{\Lambda^2({\cal H}-{\cal H}_0)\over {({\cal H}+\Lambda^2)
({\cal H}_0+\Lambda^2)}} \cr
&\qquad~~
+{k^2({\cal H}-{\cal H}_0)\over {({\cal H}+k^2)({\cal H}_0+k^2)}}\Biggr\} ,}}
which shows that $\Lambda$ may be interpreted as the mass of
some unitarity-violating ghost states. The interpretation follows from
the relative negative sign in the modified propagator.
Equivalently, one may also say that the effect of $\Lambda$
is to make the theory superrenormalizable by incorporating higher
order derivative terms. For example, eq.\cher\ implies that
the kinetic term in the scalar theory is to be modified as
\eqn\counp{ -{1\over 2}\phi\partial^2\phi\longrightarrow
{1\over
2}\phi\Bigl[-{\partial^2}+{2\over\Lambda^2}\bigl(-{\partial^2}\bigr)^2+{1\over\Lambda^4}
\bigl({-{\partial^2}}\bigr)^3\Bigr]\phi.}
On the other hand, the IR scale $k$ may be thought of as an
additional mass which makes the overall effective mass parameter
$\mu^2_{\rm eff}=\mu^2+k^2$. The scale $k$ is useful not only for the
purpose of studying RG, but can also be employed as an IR regulator for the
theory containing massless modes.

Before examining Yang-Mills theory, we consider the following
covariant fluctuation kernel:
\eqn\cdrw{ {\cal H}=-D^2+\mu^2+Y(x),}
where $D_{\mu}$ is the covariant derivative for the gauge group, $\mu^2$
the mass for the scalar field interacting with the gauge field
$A_{\mu}^a(x)$, and $Y(x)$
a matrix-valued function of $x$ describing the interaction
between the scalar particles. The index $a$ runs over the dimension of
the gauge (color) group. One may also write $Y=Y^aT^a$ where the $T^a$'s
are the generators of the gauge group satisfying
\eqn\commu{[T^a,T^b]=f^{abc}T^c,\qquad~{\rm tr}_c(T^aT^b)=
-{1\over 2}\delta^{ab}}
with $f^{abc}$ being the structure constants
and ${\rm tr_c}$ the summation over only the color indices.
In the fundamental $SU(N)$ representation, we have
\eqn\funta{ T^a=\cases{{\sigma^a/2i},&$a=1,\cdots,3$\qquad $N=2$\cr
\cr
{\lambda^a/2i},&$a=1,\cdots, 8$\qquad $N=3$,\cr}}
where $\sigma^a$ and $\lambda^a$ are, respectively, the Pauli and
the Gell-Mann matrices. When operating on $Y$ with the covariant derivative,
we have
\eqn\covar{
D^{ab}_{\mu}Y=\bigl(g^{ab}{\partial_{\mu}}-gf^{abc}A^c_{\mu}\bigr)Y,}
or $D_{\mu}Y=\partial_{\mu} Y +[A_{\mu},Y]$, where $A_{\mu}=gA^a_{\mu}T^a$
and $g$ is a coupling constant.

The unregularized one-loop contribution to the effective action is
\eqn\onfer{ \tilde S^{(1)}={1\over 2}{\rm Tr}\Bigl({\rm ln}{\cal H}-{\rm ln}
{\cal H}_0\Bigr)=-{1\over 2}\int_x\int_0^{\infty}{ds\over s}{\rm tr}\langle x|
\bigl(e^{-{\cal H} s}-e^{-{\cal H}_0s}\bigr)|x\rangle ,\qquad\int_x=\int d^dx,}
where the diagonal part of the ``heat kernel'' is written as
\eqn\diapl{\eqalign{ h(s;x,x)&=\langle x|e^{-{\cal H} s}|x\rangle =\int_p
\langle x|p\rangle e^{-{\cal H}_x s}\langle p|x\rangle
=\int_p e^{-ipx}e^{-{\cal H}_xs}e^{ipx} \cr
&
=\int_pe^{-(p^2-2ip\cdot D+{\cal H}_x)s}{\bf 1}=\int_p e^{-(p^2+\mu^2)s}
e^{(2ip\cdot D+D^2-Y)s}{\bf 1}, \qquad\int_p=\int{d^dp\over(2\pi)^d}.}}
The above expression is derived by employing the plane wave basis
$|p\rangle$ with
$\langle x|p\rangle=e^{-ipx}$ and the commutation relations
\chiral, \ref\nepomechie:
\eqn\rela{ [D_{\mu}, e^{ipx}]=ip_{\mu}, \qquad\qquad
[{\cal H}_x,e^{ipx}]=p^2-2ip\cdot D.}
The factor ${\bf 1}$ indicates that the operator $D_{\mu}$ acts on the
identity. Making use of the Baker-Campbell-Hausdorf formulae to expand the
operators in the exponential, eq.\diapl\ can be approximated as
\eqn\egrt{\eqalign{ h(s;x,x)&=e^{-(\mu^2+Y)s}\int_pe^{-p^2s}\Biggl\{
1+D^2s+{D^4\over 2}s^2-{[D^2,Y]\over 2}s^2 \cr
&
-{2p^2\over d}\Bigl[D^2s^2+{1\over
3}\Bigl([[D^2,{D_{\mu}}],{D_{\mu}}]+3{D_{\mu}}[D^2,{D_{\mu}}]
+3D^4 -[D^2,Y] \cr
&
-[{D_{\mu}},Y]{D_{\mu}}\Bigr)s^3\Bigr]+{2{(p^2)}^2\over 3d(d+2)}\Bigl[D^4
+({D_{\mu}}{D_{\nu}})^2+{D_{\mu}} D^2{D_{\mu}}\Bigr]s^4+\cdots\Biggr\}~{\bf
1},}}
where we have used the $O(d)$ invariant property of the momentum integrals:
\eqn\cvns{ \int_pp_{\mu_1}p_{\mu_2}\cdots p_{\mu_{2m}}e^{-p^2s}
={{T^m_{\mu_1\mu_2\cdots\mu_{2m}}\Gamma(d/2)}\over{2^m\Gamma(m+d/2)}}
\int_p{(p^2)}^me^{-p^2s}
={{T^m_{\mu_1\mu_2\cdots\mu_{2m}}}\over{(4\pi s)^{d/2}(2s)^m}},}
\eqn\npoer{ T^m_{\mu_1\mu_2\cdots\mu_{2m}}=\delta_{\mu_1,\mu_2}\cdots
\delta_{\mu_{2m-1},\mu_{2m}}+ {\rm~ permutations}.}
As the singularity arising from taking the spacetime trace is
transferred to the proper-time integration, we insert the regulating
smearing function ${\rho_k^{(d)}(s,\Lambda)}$ into \egrt\ and obtain the
following ``blocked''
heat kernel:
\eqn\egrst{ h_k(s;x,x)={e^{-(\mu^2+Y)s}\over
(4\pi)^{d/2}}\rho_k^{(d)}(s,\Lambda)\Biggl\{
1+{1\over 12}\Bigl[F_{\mu\nu}F_{\mu\nu}-2(D^2Y)\Bigr]s^2+O(s^3)\Biggr\},}
which is in agreement with the result found in \nepomechie\ and \ref\duff\
for ${\rho_k^{(d)}(s,\Lambda)}=1$. Gauge symmetry is easily seen to be
preserved by noting that
$h_k(s;x,x)$ consists of gauge invariant quantities only. Had we used
momentum cutoff regularization instead, there would be
contribution from noninvariant operators $D^2$, ${D_{\mu}} Y{D_{\mu}}$, $YD^2$,
$D^4$ and ${D_{\mu}} D^2{D_{\mu}}$ \ocr. Higher order contributions to \egrst\
can be included in an invariant manner as well. The details can be
found in \chiral.

\medskip
\medskip
\centerline{\bf III. YANG-MILLS THEORY}
\medskip
\nobreak
\xdef\secsym{3.}\global\meqno = 1
\medskip

In the absence of matter field the pure Yang-Mills lagrangian reads
\eqn\lagran{ {\cal L}= {1\over 4}G^a_{\mu\nu}G^a_{\mu\nu} ,}
where the field strength is given by
\eqn\fstr{G^a_{\mu\nu}={\partial_{\mu}} A^a_\nu-\partial_\nu A^a_\mu
+gf^{abc}A^b_{\mu}A^c_{\nu} ,}
or
\eqn\matfiel{ G_{\mu\nu}=gG^a_{\mu\nu}T^a={\partial_{\mu}}
A_{\nu}-{\partial_{\nu}} A_{\mu}
+[A_{\mu},A_{\nu}]}
in the matrix-valued representation. In the usual manner,
integrating out the irrelevant short-distance (fast-fluctuating) modes $\xi$
having momenta between $k$ and $\Lambda$ leads to a low-energy effective
blocked action which depends only on the
slowly-varying background fields $\bar A$ with momenta below $k$. Our
goal is to derive a gauge invariant Yang-Mills blocked action
$\tilde S_k$ which has an explicit
dependence on the IR scale $k$. This would allow us to generate an
improved RG flow equation by evolving the blocked action with $k$.
As demonstrated in Sec. II, the scale enters in an invariant manner
when operator cutoff regularization is used.

To set up the RG formalism, we first separate the modes as
\eqn\field{A^a_{\mu}(p)=\cases{\bar A^a_{\mu}(p),&$0 \le p \le k$ \cr
\cr
\xi^a_{\mu}(p), &$k < p < \Lambda$,  \cr }}
and denote the background field strength by $F^a_{\mu\nu}
=G^a_{\mu\nu}(\bar A)$. We next introduce
\eqn\gaugef{ {\cal L}_{\rm GF}= -{1\over 2\alpha}(D_{\mu}{A^a_{\mu}})^2 ,}
and
\eqn\ghost{ {\cal L}_{\rm FPG}= {\chi^{\dagger}} D^2(\bar A)\chi ,}
as the desirable gauge-fixing condition and
the Faddeev-Popov ghost term, respectively.
Notice that in order to
obtain a gauge invariant expression for the theory, the calculation is
performed in the background field formalism \ref\abb.
The lagrangian then takes on the form
\eqn\lagrann{\eqalign{ {\cal L}(\bar A_{\mu}+\xi_{\mu},
\tilde{\chi^{\dagger}}+{\eta^{\dagger}},\tilde\chi+\eta) &=
{1\over 4}F^a_{\mu\nu}F^a_{\mu\nu}+{1\over
2}{\xi^a_{\mu}}\Bigl[-D^2{g_{\mu\nu}}
+(1-{1\over \alpha}) D_{\mu}D_{\nu}\Bigr]^{ab}{\xi^b_{\nu}} \cr
&
+gf^{acb}{\xi^a_{\mu}} F^c_{\mu\nu}{\xi^b_{\nu}} +\tilde\chi^{\dagger a}
{D^2(\bar A)}^{ab}\tilde\chi^b+\eta^{\dagger a}{D^2(\bar A)}^{ab}\eta^b\cr
&
+\delta{\cal L}(\bar A_{\mu},\xi_{\mu}),}}
where ${\eta^{\dagger}}$ and $\eta$ denote the fast-fluctuating modes for the
ghost
fields, and
\eqn\dlag{ \delta{\cal L}(\bar A_{\mu},\xi_{\mu})
= gf^{abc}({D_{\mu}}\xi_{\nu})^a\xi^b_{\mu}\xi^c_{\nu}
+{1\over 4}g^2f^{abc}f^{ade}\xi^b_{\mu}\xi^c_{\nu}\xi^d_{\mu}\xi^e_{\nu}
+\cdots }
represents the higher-order self-interactions.
The partition function can be written as
\eqn\part{Z=\int{\cal D}[A_\mu]{\cal D}[\chi]
{\cal D}[\chi^\dagger]e^{-S
[\bar A_\mu+\xi_\mu,\tilde{\chi^{\dagger}}+{\eta^{\dagger}},\tilde\chi+\eta]}
=\int{\cal D}[{\bar A}]{\cal D}[\tilde\chi]{\cal D}[\tilde{\chi^{\dagger}}]
e^{-\tilde S_k[{\bar A}_{\mu},\tilde{\chi^{\dagger}},\tilde\chi]},}
where
\eqn\blint{e^{-\tilde S_k[\bar A_{\mu},\tilde{\chi^{\dagger}},\tilde\chi]}=
\int{\cal D}[\xi_\mu]{\cal D}[\eta]{\cal D}
[\eta^\dagger]e^{-S[\bar A_\mu+\xi_{\mu},\tilde{\chi^{\dagger}}
+{\eta^{\dagger}},\tilde\chi+\eta]}.}
In the above, the functional integrations are performed within the perspective
momentum range for each field configuration.
In the case of vanishing ghost background fields,
by substituting \lagrann\ into \blint\ and dropping higher order fluctuating
terms, the operator cutoff regularized blocked action up to the one-loop
order reads
\eqn\newl{\eqalign{\tilde S_k[\bar A] &={1\over 4}\int_xF^a_{\mu\nu}
F^a_{\mu\nu}+{1\over2}{\rm Tr}_{\rm oc}\Bigl[{\rm ln}{\cal K}({\bar A})
-{\rm ln}{\cal K}(0)\Bigr]-{\rm Tr}_{\rm oc}\Bigl[{\rm ln}{\cal O}({\bar A})
-{\rm ln}{\cal O}(0)\Bigr],}}
where the gauge and the ghost kernels are, respectively,
\eqn\kerne{ {\cal K}^{ab}_{\mu\nu}={{\partial^2 S}\over{\partial A^a_{\mu}(x)
\partial A^b_{\nu}(y)}}\Big\vert_{{\bar A}}=\Biggl\{-\Bigl[D^2{g_{\mu\nu}}
-(1-{1\over \alpha})
D_{\mu}D_{\nu}\Bigr]^{ab}+2gf^{abc}F^c_{\mu\nu}\Biggr\}\delta^4(x-y),}
and
\eqn\kerno{ {\cal O}^{ab}=-{D^2(\bar A)}^{ab}\delta^4(x-y) .}
Eq.\kerne\ is derived by the help of
\eqn\comm{ D^{ab}_{\mu}D^{bc}_{\nu}-D^{ab}_{\nu}D^{bc}_{\mu}
=gf^{abc}F^b_{\mu\nu}.}
Here ${\rm Tr}_{\rm oc}$ denotes the trace sum over (operator cutoff
regularized) space-time, Lorentz
indices as well as the color indices and ${\rm tr}$ is for the
latter two only. When no confusion arises, internal indices
shall be suppressed for brevity.

Since the Yang-Mills blocked action is generally a complicated object even
at the one-loop elvel, an approximate solution exists only in a
certain energy regime. We shall follow the perturbative formalism developed by
Schwinger in \schwinger.
In the momentum space where ${\partial_{\mu}}\to ip_{\mu}$,
\kerne\ and \kerno\ may be rewritten as \ref\mckeon
\eqn\kernee{\eqalign{ {\cal K}^{ab}_{\mu\nu}&=-\Bigl[D^2{g_{\mu\nu}}
-(1-{1\over\alpha})D_{\mu}D_{\nu}\Bigr]^{ab}+2gf^{abc}F^c_{\mu\nu} \cr
&
={g_{\mu\nu}}\Bigl[p^2\delta^{ab}-igf^{acb}(p\cdot {\bar A}^c
+{\bar A}^c\cdot p)-g^2f^{amc}f^{c\ell b}{\bar A}^m_{\lambda}
{\bar A}^{\ell}_{\lambda}\Bigr] +2gf^{abc}F^c_{\mu\nu} \cr
&
-(1-{1\over\alpha})\Bigl[p_{\mu}p_{\nu}{\delta^{ab}}-igf^{acb}(p_{\mu}{\bar
A}_{\nu}^c
+{\bar A}_{\mu}^c p_{\nu})-g^2f^{amc}f^{c\ell b}{\bar A}^m_{\mu}
{\bar A}^{\ell}_{\nu}\Bigr] ,}}
and
\eqn\gker{ {\cal O}^{ab}=-{(D^2)}^{ab}=p^2\delta^{ab}-igf^{acb}(p\cdot {\bar
A}^c
+{\bar A}^c\cdot p)-g^2f^{amc}f^{c\ell b}{\bar A}^m_{\lambda}
{\bar A}^{\ell}_{\lambda} .}
With ${\cal H}={\cal H}_0
+{\cal H}_I$ where ${\cal H}_I$ accounts for the interactions, we
and expand the fluctuation kernel in power of $s$ and obtain
\eqn\expan{\eqalign{{\rm Tr}\bigl(e^{-{\cal H}s}\bigr)&={\rm Tr}\bigl(
e^{-{\cal H}_0s}\bigr)+\int_0^{\infty}d\lambda~{\rm Tr}\Bigl({\cal H}_I
e^{-({\cal H}_0+\lambda{\cal H}_I)s}\Bigr) \cr
&
={\rm Tr}\Biggl\{e^{-{\cal H}_0s}+(-s)e^{-{\cal H}_0s}{\cal H}_I+{(-s)^2\over
2}
\int_0^1du_1e^{-(1-u_1){\cal H}_0s}{\cal H}_Ie^{-u_1{\cal H}_0s}{\cal H}_I \cr
&
+{(-s)^3\over 3}\int_0^1du_1u_1\int_0^1du_2e^{-(1-u_1){\cal H}_0s}{\cal H}_I
e^{-u_1(1-u_2){\cal H}_0s}{\cal H}_Ie^{-u_1u_2{\cal H}_0s}{\cal
H}_I+\cdots\Biggr\}.}}
Applying the above expansion formula to \kernee\ and \gker, we have
\eqn\gho{\eqalign{ {\rm Tr}\bigl(e^{{-{\cal O}}^{ab}s}\bigr)
&={\rm Tr}~\delta^{ab}\Biggl\{e^{-p^2s}-se^{-p^2s}\bigl(-g^2f^{amc}f^{c\ell b}
{\bar A}^m_{\lambda}{\bar A}^{\ell}_{\lambda}\bigr) \cr
&
+{s^2\over 2}\int_0^1du e^{(1-u)p^2s}\bigl[-igf^{ac\ell}(p\cdot{\bar A}^c+
{\bar A}^c\cdot p)\bigr] \cr
&\qquad
\times e^{-up^2s}\bigl[-igf^{\ell mb}(p\cdot{\bar A}^m
+{\bar A}^m\cdot p)\bigr]+\cdots\Biggr\} ,}}
and, in Feynman gauge where $\alpha=1$,
\eqn\deed{\eqalign{ {\rm Tr}\bigl(e^{{-{\cal K}}^{ab}_{\mu\nu}s}\bigr)
&={\rm Tr}~{\delta^{ab}}\Biggl\{{g_{\mu\nu}} e^{ -p^2s}-se^{-p^2s}
\Bigl(-g^2f^{amc}f^{c\ell b}{g_{\mu\nu}}
{\bar A}^m_{\lambda}{\bar A}^{\ell}_{\lambda}\Bigr) \cr
&
+{s^2\over 2}\int_0^1du e^{(1-u)p^2s}\Bigl[-igf^{ac\ell}g_{\mu\rho}
(p\cdot{\bar A}^c+{\bar A}^c\cdot p)-2gf^{ac\ell}F^c_{\mu\rho}({\bar A})\Bigr]
\cr
&\qquad
\times e^{-up^2s}\Bigl[-igf^{\ell mb}g_{\rho\nu}(p\cdot{\bar A}^m
+{\bar A}^m\cdot p)-2gf^{\ell mb}F^m_{\rho\nu}({\bar A})\Bigr]+\cdots\Biggr\}
.}}
By inserting a complete orthonormal set of momentum
states $|p\rangle$ satisfying:
\eqn\renor{\eqalign{\int_p{|p\rangle\langle p|}&=1,
\quad {\langle p|p'\rangle}=(2\pi)^4\delta^4(p-p') \cr
\cr
{\langle x|p\rangle}&=e^{i{p\cdot x}},\quad {\langle p|A_{\mu}|p'\rangle}
=A_{\mu}(p-p') ,}}
we carry out the calculations explicitly in $d=4$ and obtain
\eqn\seco{\eqalign{& {\rm Tr}\int_0^1du~e^{-(1-u)p^2s}
(p\cdot{\bar A}^c+{\bar A}^c\cdot p)e^{-up^2s}(p\cdot{\bar A}^c+{\bar A}^c\cdot
p) \cr
&
={\rm tr}\int_0^1du\int_{p,q} e^{-[(1-u)p^2+uq^2]s}(p+q)_{\mu}(p+q)_{\nu}
{\bar A}^c_{\mu}(p-q){\bar A}^c_{\nu}(q-p) \cr
&
={1\over 16\pi^2s^2}~{\rm tr}\int_0^1du\int_p{\bar A}^c_{\mu}(p){\bar
A}^c_{\nu}(-p)
e^{-u(1-u)p^2s}\Bigl[{2{g_{\mu\nu}}\over s}+(2u-1)^2{p_{\mu}p_{\nu}}\Bigr] .}}
The above expression is arrived by shifting the variable $p\to p+q$
followed by $q\to q-(1-u)p$, and the $q$ integration using $O(4)$ invariance.
Making use of the regulating smearing function $\rho_k^{(4)}(s,\Lambda)$ then
leads to
\eqn\tde{\eqalign{{\rm Tr}_{\rm oc}{\rm ln}{\cal O}({\bar A})&=
{g^2C_2(G)\over 16\pi^2}\int_p{\bar A}^c_{\mu}(p){\bar A}^c_{\nu}(-p)
\Biggl\{{g_{\mu\nu}}\int_0^{\infty}{ds\over s^2}\rho_k^{(4)}(s,\Lambda) \cr
&
-{1\over 2}\int_0^1du\int_0^{\infty}{ds\over s}\rho_k^{(4)}(s,\Lambda)
e^{-u(1-u)p^2s}\Bigl[{2{g_{\mu\nu}}\over s}+(2u-1)^2{p_{\mu}p_{\nu}}\Bigr]
\Biggr\}\cr
&
={g^2C_2(G)\over 32\pi^2}\int_p{\bar A}^c_{\mu}(p){\bar A}^c_{\nu}(-p)
\int_0^1du\Biggl\{{p_{\mu}p_{\nu}}(2u-1)^2\Bigl[{\tilde\Lambda^2\over
{\tilde\Lambda^2+u(1-u)}}\cr
&
-{\kappa^2\over {\kappa^2+u(1-u)}}\Bigr]
-\Bigl[2{g_{\mu\nu}} p^2u(1-u)-(2u-1)^2{p_{\mu}p_{\nu}}\Bigr]{\rm
ln}\Bigl({{\kappa^2+u(1-u)}\over
{\tilde\Lambda^2+u(1-u)}}\Bigr) \Biggr\} \cr
&
={g^2C_2(G)\over 192\pi^2}\int_p F^c_{\mu\nu}(p)F^c_{\mu\nu}(-p)
\Biggl\{ {\rm ln}{\Lambda^2\over k^2}+{5\over 3}
-4\kappa^2-(2\kappa^2-1)(4\kappa^2+1)f(\kappa)\Biggr\} ,}}
where $\kappa=k/p$ and
\eqn\flo{ f(\kappa)={1\over {\sqrt{4\kappa^2+1}}}~{{\rm
ln}\Bigl({{\sqrt{4\kappa^2+1}-1}\over{\sqrt{4\kappa^2+1}+1}}
\Bigr)} .}
In the above, we have made the substitution
\eqn\ker{ {1\over
2}F^c_{\mu\nu}(p)F^c_{\mu\nu}(-p)=(p^2{g_{\mu\nu}}-{p_{\mu}p_{\nu}})
{\bar A}^c_{\mu}(p){\bar A}^c_{\nu}(-p) +\cdots,}
by neglecting higher order gauge field contributions, as well as
the expansion
\eqn\appr{ -{1\over 2}{\rm ln}\Bigl({1-x\over {1+x}}\Bigr)
={\rm tanh}^{-1}x=x+{x^3\over 3}+\cdots,}
by taking the limit $\Lambda\to\infty$.
Notice that $f^{ac\ell}f^{\ell ma}=-\delta^{cm}C_2(G)$,
where $C_2(G)$ is a Casimir operator with $C_2(G)=N$ for $G=SU(N)$.
The quadratic divergence naively expected from dimensional counting also
disappears, as required by gauge invariance.

Similarly, the gauge field contribution reads
\eqn\secon{\eqalign{ {\rm Tr}_{\rm oc}{\rm ln}{\cal K} &({\bar A})=
{g^2C_2(G)\over 4\pi^2}\int_p{\bar A}^c_{\mu}(p){\bar A}^c_{\nu}(-p)
{g_{\mu\nu}}\int_0^{\infty}{ds\over s^2}\rho_k^{(4)}(s,\Lambda) \cr
&
-{1\over 2}{\rm Tr}\int_0^{\infty}ds~ s\rho_k^{(4)}(s,\Lambda)
\Biggl(\int_0^1du~
e^{-(1-u)p^2s}\delta^{ab}\Bigl\{-igf^{ac\ell}\bigl[(p\cdot{\bar A}^c
+{\bar A}^c\cdot p)g_{\mu\rho} \cr
&
-2iF^c_{\mu\rho}\bigr]\Bigr\}e^{-up^2s}\Bigl\{-igf^{\ell mb}
\bigl[(p\cdot{\bar A}^m+{\bar A}^m\cdot p)
g_{\rho\nu}-2iF^m_{\rho\nu}\bigr]\Bigr\} \Biggr) \cr
&
= 4{\rm Tr}_{\rm oc}{\rm ln}{\cal O}({\bar A})-{g^2C_2(G)\over 8\pi^2}
\int_p F^c_{\mu\rho}(p)F^c_{\mu\rho}(-p)
\int_0^1du\int_0^{\infty}{ds\over s}\rho_k^{(4)}(s,\Lambda)e^{-u(1-u)p^2s} \cr
&
=4{\rm Tr}_{\rm oc}{\rm ln}{\cal O}({\bar A})-{g^2C_2(G)\over 8\pi^2}\int_p
F^c_{\mu\rho}(p)F^c_{\mu\rho}(-p)
\Biggl\{ {\rm ln}\bigl({\Lambda^2\over k^2}\bigr)+1+(2\kappa^2+1)f(\kappa)
\Biggr\}.}}
Adding up these terms, the perturbative Yang-Mills blocked action becomes
\eqn\ymbl{\tilde S_k ={1\over 4}\int_p F^a_{\mu\nu}
F^a_{\mu\nu}+{1\over2}{\rm Tr}_{\rm oc}{\rm ln}{\cal K}(\bar A)
-{\rm Tr}_{\rm oc}{\rm ln}{\cal O}(\bar A)
= {1\over 4}\int_p\tilde {\cal Z}_k^{-1}(p)F^a_{\mu\nu}(p)F^a_{\mu\nu}(-p),}
where
\eqn\zzz{\eqalign{\tilde{\cal Z}_k^{-1}&=1-{g^2C_2(G)\over 48\pi^2}
\Biggl\{11~{\rm ln}\Bigl({\Lambda^2\over k^2}\Bigr)
+{31\over 3}+4\kappa^2+\bigl(8\kappa^4+22\kappa^2+11\bigr)f(\kappa)\Biggr\}\cr
&
={\cal Z}_k^{-1}+\delta\tilde{\cal Z}_k^{-1},}}
with
\eqn\zzr{{\cal Z}_k^{-1}= 1-{g^2C_2(G)\over 48\pi^2}\Bigl[11{\rm ln}
\Bigl({\Lambda^2\over k^2}\Bigr)+{1\over 3}\Bigr]
=1-{11g^2C_2(G)\over 48\pi^2}{\rm ln}\Bigl({\tilde\Lambda^2\over k^2}\Bigr),}
and
\eqn\zzo{\delta\tilde{\cal Z}_k^{-1}= -{g^2C_2(G)\over 48\pi^2}\Bigl[
{10\over 3}+4\kappa^2+\bigl(8\kappa^4+22\kappa^2+11\bigr)f(\kappa)\Bigr].}
Here we recover the familiar factor of $-11g^2C_2(G)/{48\pi^2}$ associated
with the ${\rm ln}\Lambda^2$ term, with $-10g^2C_2(G)/{48\pi^2}$ coming
from the gauge kernel ${\rm Tr}_{\rm oc}{\rm ln}{\cal K}/2$ and
$-g^2C_2(G)/{48\pi^2}$ from the ghost sector
$-{\rm Tr}_{\rm oc}{\rm ln}{\cal O}$. The constant $1/3$ inside
the bracket of \zzr\ is regularization scheme dependent and can be adjusted by
changing the cutoff $\Lambda\to\tilde\Lambda$, i.e. a finite renormalization.

However, eq.\ymbl\ is not gauge invariant since the complicated
non-polynomial $p$ dependence coming from $\delta\tilde{\cal Z}_k^{-1}$
implies
the presence of nonlocal field strength coupling
which manifestly violates gauge symmetry. Although nonlocality may be
characteristic of momentum cutoff brought about by the regulating smearing
function, this is a general feature for the low energy blocked action
irrespective of how the theory is regularized.
Nevertheless, the difficulty can be avoided by taking
the large $k$ limit where
\eqn\zse{\delta\tilde{\cal Z}_k^{-1}\approx {g^2C_2(G)\over 288\pi^2}
\Bigl\{ 7{p^2\over k^2}+{1\over 6}{\bigl({p^2\over k^2}\bigr)}^2
+\cdots\Bigr\}.}
Notice that as $k\to\infty$, $\delta\tilde{\cal Z}_k^{-1}\to 0$
and the theory is completely local.
The expansion enables us to ``upgrade'' the momentum $p$ to the
generalized momentum $\Pi$ (up to terms which are of higher order in ${\bar
A}$)
as
\eqn\upgr{\eqalign{ \int_pF^c_{\mu\nu}(p)p^2F^c_{\mu\nu}(-p)
&\longrightarrow \int_p F^c_{\mu\nu}(p)\Pi^2F^c_{\mu\nu}(-p)\longrightarrow
\int_x D_{\mu}F^c_{\mu\nu}(x)D_{\sigma}F^c_{\sigma\nu}(x) \cr
\int_pF^c_{\mu\nu}(p){(p^2)}^2F^c_{\mu\nu}(-p)
&\longrightarrow  \int_pF^c_{\mu\nu}(p){(\Pi^2)}^2F^c_{\mu\nu}(-p)
\longrightarrow \int_xD^2F^c_{\mu\nu}(x)D^2F^c_{\mu\nu}(x) ,}}
which leads to the following gauge invariant blocked potential:
\eqn\elk{U_k={{\cal Z}_k^{-1}\over 4}F^a_{\mu\nu}F^a_{\mu\nu}
+{g^2C_2(G)\over 1152\pi^2}\Biggl\{
{7\over k^2}\bigl(D_{\mu}F^a_{\mu\nu}D_{\sigma}F^a_{\sigma\nu}\bigr)
+{1\over 6k^4}\bigl(D^2F^a_{\mu\nu}D^2F^a_{\mu\nu}\bigr)+\cdots\Biggr\}.}
These higher dimensional
operators which are completely absent at the cutoff scale $k=\Lambda$
may be thought of as being generated by blocking
transformation as $k$ is lowered. While the
presence of logarithmic divergence in ${\cal Z}^{-1}_k$ can be
readily absorbed
by a redefinition of the coupling constant $g$ associated with the operator
$F_{\mu\nu}^aF^a_{\mu\nu}$, the contribution from $\delta\tilde{\cal Z}_k^{-1}$
accounts for the finite renormalization of the coupling strengths for the
higher dimensional operators.
Notice that the coefficients for the higher operators generated from
the above heuristic argument are expected to differ from the formal
covariant derivative expansion. A thorough treatment on this subject may be
found in \chiral.

\bigskip
\medskip
\centerline{\bf IV. RENORMALIZATION GROUP EQUATION}
\medskip
\nobreak
\xdef\secsym{4.}\global\meqno = 1
\medskip

We now examine the RG flow pattern of
the Yang-Mills blocked action derived in the last Section.
Using Slavnov-Taylor identities,
the field renormalization constant ${\cal Z}_k$ written in
\zzz\ can be related to the coupling constant renormalization by
$g^2={\cal Z}_k^{-1}g^2(k)$ where
\eqn\coupl{ {1\over g^2(k)}={1\over g^2}-{C_2(G)\over 48\pi^2}\Bigl[
11{\rm ln}\Bigl({\Lambda^2\over k^2}\Bigr)+{1\over 3}\Bigr]
={1\over g^2}-{11C_2(G)\over 48\pi^2}
{\rm ln}\Bigl({\tilde\Lambda^2\over k^2}\Bigr).}
The running coupling constant $g(k)$ reduces to the usual bare coupling
$g$ at $k=\tilde\Lambda$, as expected from perturbation theroy. However, in the
low energy regime where $k\to 0$, IR singularity appears due to the
masslessness of the gluon fields. The renormalized coupling constant is
usually defined at an arbitrary off-shell scale $k_0$ as
\eqn\fed{ g^2_R(k_0)=g^2(k^2=k_0^2)={g^2\over {1-
{11C_2(G)g^2\over 48\pi^2}{\rm ln}\Bigl({\tilde\Lambda^2\over k_0^2}\Bigr)}}.}
Similarly, the $\beta$ function reads
\eqn\btafun{\beta(g(k))=k{{\partial g(k)}\over{\partial k}}=
-{11C_2(G)\over 48\pi^2}g^3(k). }
On the other hand, the evolution of the blocked action is obtained by
differentiating \ymbl\ with respect to $k$ to give
\eqn\floo{ k\partial_k\tilde S_k=
{1\over 4} k{{\partial{\cal Z}_k^{-1}\over{\partial k}}\int_p
F^a_{\mu\nu}(p)F^a_{\mu\nu}(-p),}}
which along with \zzr\ and \coupl, implies that
this ''independent-mode'' approximation \ref\mike\ simply reduces
\floo\ to the $\beta$ function \btafun\ which governs the flow of the
coupling constant. The result is to be expected on the ground that
$g$ is the only free parameter
in the theory. It also justifies the truncation of the background
fields beyond quadratic order in our
perturbative evaluation of the Yang-Mills blocked
action.

While perturbation works for large $k$ where the theory exhibits
asymptotic freedom, it breaks down in the low
energy limit as the coupling constant grows stronger.
Thus, for small $k$, in addition to ${\cal Z}_k^{-1}$, one must
also consider the nonlocal sector $\delta\tilde{\cal Z}_k^{-1}$,
for nonlocal interactions may be a crucial ingredient for explaining
confinement. Hence, from \zzz, we have the following
nonlocal running coupling constant (denoted with a tilde):
\eqn\cout{ {1\over {\tilde g}^2(\kappa)}={1\over g^2}-{C_2(G)\over 48\pi^2}
\Bigl[11~{\rm ln}\Bigl({\Lambda^2\over k^2}\Bigr)
+{31\over 3}+4\kappa^2+(8\kappa^4+22\kappa^2+11)f(\kappa)\Bigl],}
which gives
\eqn\bfu{\tilde\beta({\tilde g}(\kappa),\kappa)=\kappa{{\partial{\tilde g}
(\kappa)}\over{\partial\kappa}}
={C_2(G)\over 8\pi^2}{\tilde g}^3(\kappa){\kappa^2\over {4\kappa^2+1}}\Bigl\{
-3+4\kappa^2+2\kappa^2(5+4\kappa^2)f(\kappa)\Bigr\}.}
In other words, the behavior of the coupling constant in the IR region
is characterized by the complicated expression in \bfu.
Notice that as $k\to\Lambda$, \cout\ and \bfu\ reduce to
the usual perturbative expressions found in \coupl\ and \btafun, respectively.

Besides nonlocal interactions, higher dimensional operators
may also play an important role in the physics of confinement in spite of
their initial suppression at the high energy scale.
The RG equation in \bfu\ was obtained by an expansion in
$s$ up to $O(s^2)$;
contributions from higher order operators such as $(F^2)^2$ are
therefore neglected.
To improve the evolution equation \floo\ in the spirit of Wilson-Kadanoff,
we first turn to the simplest noninvariant momentum cutoff regularization since
its main features may be reproduced in the invariant operator cutoff
scheme. From \newl, we have
\eqn\neww{ k\partial {\tilde S}_k={1\over 2}k\partial_k\Biggl\{{\rm Tr'}
\Bigl[{\rm ln}{\cal K}_{\mu\nu}^{ab}({\bar A})-{\rm ln}{\cal
K}_{\mu\nu}^{ab}(0)
\Bigr]-2{\rm Tr'}\Bigl[{\rm ln}{\cal O}^{ab}({\bar A})-{\rm ln}{\cal O}^{ab}(0)
\Bigr]\Biggr\},}
where the prime notation in ${\rm Tr'}$ indicates the presence of cutoff scale
in the momentum integration. In going beyond the one-loop approximation to
probe the
physics near the energy scale $\sim k$ using the RG improved idea,
the first step is to divide the momentum integration volume defined
between $k$ and the UV cutoff $\Lambda$
into a large number of thin shells each having a width $\Delta k$. By lowering
the cutoff infinitesimally from $\Lambda\to\Lambda-\Delta k\to\Lambda
-2\Delta k$ until the scale $k$ is reached, we then arrive at a nonlinear
partial differential RG equation which takes into account the continuous
feedbacks of the high mode to the lower ones as it gets integrated over\mike.
Following this
prescription, we obtain the following RG improved equation:
\eqn\newr{ k\partial {\tilde S}_k={1\over 2}k\partial_k\Biggl\{{\rm Tr'}
\Bigl[{\rm ln}\Bigl({{\partial^2\tilde S_k}\over{\partial A_{\mu}^a\partial
A_{\nu}^b}}\Bigr)_{{\bar A}}-{\rm ln}\Bigl({{\partial^2\tilde
S_k}\over{\partial
A_{\mu}^a\partial A_{\nu}^b}}\Bigr)_0\Bigr]-2{\rm Tr'}\Bigl[
{\rm ln}{\cal O}^{ab}({\bar A})-{\rm ln}{\cal O}^{ab}(0)\Bigr]\Biggr\},}
which is similar to that derived in \rt.
Notice that there is no ``dressing'' in the
ghost sector since the ghost fields enter the action as an ``initial''
boundary condition.
By comparing \newr\ with \neww, we observe that the ``trick''
to implement RG seems to be a simple replacement of the bare action $S$
in the definition of
${\cal K}_{\mu\nu}^{ab}$ by its corresponding $k$-dependent blocked action
$\tilde S_k$. This dressing is equivalent to summation over the
higher order nonoverlapping
graphs such as the daisies and the superdaisies when investigating theory
at finite temperature \mike.

Alternatively, we may first
go back to \neww\ and split the fluctuation kernels as
\eqn\sep{ {\cal K}_{\mu\nu}^{ab}({\bar A})={\cal K}_{\mu\nu}^{ab}(0)
+\delta{\cal K}_{\mu\nu}^{ab}({\bar A}),\qquad
{\cal O}^{ab}({\bar A})={\cal O}^{ab}(0)+\delta{\cal O}^{ab}({\bar A}),}
where ${\cal K}^{ab}_{\mu\nu,0}={\hat 1}^{ab}_{\mu\nu}p^2$ and
${\cal O}^{ab}_0=\delta^{ab}p^2$ with ${\hat 1}^{ab}_{\mu\nu}$
being the unit matrix in Lorentz and color space. In case where
$\delta\cal K$ and $\delta\cal O$ are constant, instead of the
blocked action, it suffices to
consider the blocked potential whose evolution equation is given by
\eqn\evk{k\partial_kU_k=
-{k^4\over16\pi^2}{\rm tr}{\rm ln}
\Bigl({{{\hat 1}^{ab}_{\mu\nu}k^2+\delta{\cal K}^{ab}_{\mu\nu}({\bar A})} \over
{{\hat 1}^{ab}_{\mu\nu}k^2}}\Bigr)+{k^4\over 8\pi^2}{\rm tr}{\rm ln}
\Bigl({{\delta^{ab}k^2+\delta{\cal O}^{ab}({\bar A})}\over
\delta^{ab}k^2}\Bigr),}
subject to the boundary condition
\eqn\uvbc{\lim_{k\to\Lambda}U_k=-{1\over2g^2}{\rm tr_c}\bigl(F_{\mu\nu}
F_{\mu\nu}\bigr)\Big\vert_{k=\Lambda}
={1\over 4}F^a_{\mu\nu}F^a_{\mu\nu}\Big\vert_{k=\Lambda}={1\over 4}
G^a_{\mu\nu}G^a_{\mu\nu}={\cal L},}
i.e., the bare lagrangian is recovered at the UV cutoff scale.
Summing over the Lorentz and color indices, \evk\ simplifies to
\eqn\evk{k\partial_kU_k=-{k^4\over16\pi^2}{\rm ln}
\Bigl({{k^2+\delta{\cal K}({\bar A})} \over k^2}\Bigr)+{k^4\over 8\pi^2}{\rm
ln}
\Bigl({{k^2+\delta{\cal O}({\bar A})}\over k^2}\Bigr),}
which implies the following improved equation:
\eqn\evk{k\partial_kU_k=-{k^4\over16\pi^2}{\rm ln}
\Bigl({{k^2+U_k''} \over k^2}\Bigr)+{k^4\over 8\pi^2}{\rm ln}
\Bigl({{k^2+\delta{\cal O}({\bar A})}\over k^2}\Bigr).}
Eq. \evk\ is applicable in the case of homogeneous background such as
constant magnetic or chromomagnetic field which we explore in the next
Section.

Going to the invariant operator cutoff prescription, \newr\ takes on the form
\eqn\eve{\eqalign{k\partial_k \tilde S_k&=-{1\over 2}~{\rm Tr}
\int^{\infty}_0{ds\over s}
k{\partial\rho_k^{(4)}(s,\Lambda)\over {\partial k}}\Biggl\{
\Bigl[e^{-{\cal K}^{ab}_{\mu\nu,k}({\bar A})s}-e^{-{\cal
K}^{ab}_{\mu\nu}(0)s}\Bigr] \cr
&\qquad\qquad
-2\Bigl[e^{-{\cal O}^{ab}({\bar A})s}-e^{-{\cal O}^{ab}(0)s}\Bigr]\Biggr\},}}
where
\eqn\kerrr{{\cal K}^{ab}_{\mu\nu,k}({\bar A})={{\partial^2\tilde S_k}\over
{\partial A^a_{\mu}\partial A^b_{\nu}}}\Big\vert_{{\bar A}}.}
Again we see that the improved RG takes on the form of a nonlinear partial
differential equation. If
\eve\ is approximated by expanding the integrand in power of $s$
and keeping only up to $O({\bar A}^2)$, the flow would
reduce to the usual $\beta$-function, as we have seen before.
Since treating $s$ as a small expansion
parameter corresponds to
exploring the high energy regime of the theory, it is not surprising
after all
that the short-distance property of asymptotic freedom is easily recovered from
such a perturbative approximation of the blocked action.
However, if one is interested in
the IR behavior of the theory, the approximation becomes unreliable.
Actually a complete $s$ integration without expansion is
possible and it gives
\eqn\evy{\eqalign{ k\partial_k \tilde S_k=
k^4~&{\rm Tr}\Biggl\{ \Bigl[{\hat 1}^{ab}_{\mu\nu}k^2
+{\cal K}^{ab}_{\mu\nu,k}({\bar A})\Bigr]^{-2}-
\Bigl[{\hat 1}^{ab}_{\mu\nu}k^2+{\cal K}^{ab}_{\mu\nu,k}(0)\Bigr]^{-2} \cr
&
-2\Bigl[ \delta^{ab}k^2+{\cal O}^{ab}({\bar A})\Bigr]^{-2}+
2\Bigl[\delta^{ab} k^2+{\cal O}^{ab}(0)\Bigr]^{-2} \Biggr\} .}}
The role played by higher
dimensional operators at the energy scale $\sim k$ can now be elucidated by
solving \evy. The above equation may be compared with \newr\
which is obtained using
the noninvariant momentum cutoff regularization. The resulting effective
blocked actions generated from these nonlinear partial differential equations
are expected to be nonlocal. Nevertheless, if the relevant local
operators, be they of higher dimension or not, can be identified to account
for the physical phenomena in the infrared, a
suitable expansion may then be possible to make the theory local.

\medskip
\goodbreak
\medskip
\centerline{\bf V. CONSTANT CHROMOMAGNETIC FIELD}
\xdef\secsym{5.}\global\meqno = 1
\medskip

Enormous efforts have been devoted to the study of the vacuum structure
of $SU(2)$ gauge theory in a constant chromomagnetic background since the
pioneering work of Matinyan and Savvidy \ref\matinyan\ -[25].
By the help of \evy,
we now explore the RG evolution associated with this configuration.

For simplicity, we choose the background to be a constant chromomagnetic
field $B$ in the $\hat z$-direction produced by
\eqn\bac{ {\bar A}^a_{\mu}=\delta^{a3}\delta_{\mu 2}Bx,}
with
\eqn\stren{ F^a_{\mu\nu}F^a_{\mu\nu}=2B^2.}
An alternative choice ${\bar A}^a_{\mu}={1\over 2}B\delta^{a3}(x\delta_{\mu 2}
-y\delta_{\mu 1})$ has also been used in \ref\levi.
Working in the background gauge, the eigenvalues for the kernels
${\cal K}$ and ${\cal O}$ can be obtained by a diagonalization in the color
space, which then reduces the equation of motion into a harmonic oscillator
equation and yields the Landau energy levels labeled by
$n$, where $n=0,1,2,\cdots$ \ref\maiani\ \ref\nielsen.
Thus, the Yang-Mills blocked potential can be written
as \ref\olsen
\eqn\poerf{ U_k \sim \int {dp\over 2\pi}\sum_{n=0}^{\infty}\sum_{S_z=\pm 1}
\sqrt{p^2+k^2+2gB(n+{1\over 2})-2gBS_z},}
where $S_z$ is the $\hat z$-component of the gluon spin along
the direction of the
chromomagnetic field, and the factor 2 in $\vec B\cdot\vec S$
comes from the gyromagnetic ratio
$g_{\rm L}=2$ for the gluon fields. Since gluons are massless vector
particles, $S_z=0$ is an
unphysical degree of freedom and the associated contribution will be
cancelled by the Faddeev-Popov ghost \olsen. For
$n=0$ and $S_z=1$, we notice that $U_k$ becomes complex
below certain momentum scale.
The unstable mode gives an
imaginary contribution to the blocked potential and signals an instability
for the vacuum. Thus, to stabilize the theory, we choose the IR
scale $k$ to be such that $k^2 > gB$.

Using
\eqn\crte{ \int{dp\over 2\pi}\sqrt{p^2+E^2}=\int{d^2p\over (2\pi)^2}{\rm ln}
\bigl(p^2+E^2\bigr),}
up to an $E$-independent constant, the trace sum in the operator cutoff
formalism may be represented as
\eqn\trace{ {\rm Tr}_{\rm oc}=\Omega {gB\over 2\pi}
\rho_k^{(2)}(s,\Lambda)\int{d^2p\over (2\pi)^2}\sum_n ,}
where  $\Omega$ is the space-time volume. Taking
into account the multiplicity factors for the eigenvalues,
it follows from \evy\ that the RG equation for the theory reads
\eqn\evvy{\eqalign{ k&\partial_k U_k ={k^2gB\over 2\pi}~2
\int{d^2p\over (2\pi)^2} \Biggl\{ \Bigl[{1\over {p^2+k^2-gB}}
-{1\over {p^2+k^2}} \Bigr]
+3\Bigl[{1\over {p^2+k^2+gB}}
-{1\over {p^2+k^2}}\Bigr] \cr
&
+4\sum_{n=1}^{\infty}\Bigl[{1\over {p^2+k^2+(2n+1)gB}}-
{1\over {p^2+k^2}}\Bigr]
-2\sum_{n=0}^{\infty}\Bigl[{1\over {p^2+k^2+(2n+1)gB}}-
{1\over {p^2+k^2}}\Bigr]\Biggr\} \cr
&
=-{k^2gB\over 4\pi^2}\Biggl\{ {\rm ln}\Bigl({{k^2-gB}\over k^2}\Bigr)
+3{\rm ln}\Bigl({{k^2+gB}\over k^2}\Bigr)
+4\sum_{n=1}^{\infty}{\rm ln}\Bigl({{k^2+(2n+1)gB}\over k^2}\Bigr)\cr
&\qquad\qquad
-2\sum_{n=0}^{\infty}{\rm ln}\Bigl({{k^2+(2n+1)gB}\over k^2}\Bigr)\Biggr\},}}
where the overall factor of 2 accounts
for the color charge degeneracy in the $SU(2)$ gauge group.
While the last term in the curly bracket represents the contribution from
the Faddeev-Popov ghost kernel, the first term is due to the
mode which becomes unstable for $k^2< gB$.
Notice that the multiplicity factors for the eigenvalues
$gB$ and $(2n+1)gB$ for $n>1$ were miswritten in \maiani; the correct factors
should be 3 and 4, respectively. The reason is due
to the negligence of the unphysical $S_z=0$ sector which yields
eigenvalues $(2n+1)gB$ for $n=0,1,\cdots$. As explained before,
this mode must be
considered fully in the presence of Faddeev-Popov ghost.

With $g(k)={\cal Z}^{1/2}g$ and
\eqn\zze{ {\cal Z}_k^{-1}={{\partial U_k}\over\partial{\cal F}}
={1\over B}{{\partial U_k}\over\partial B},}
the $\beta$ function can be rewritten as
\eqn\bbt{\eqalign{\beta &(g(k),\tau)=k\partial_k g(k)=-{1\over 2}g
{\cal Z}_k^{3/2}k\partial_k {\cal Z}_k^{-1}
=-g{\cal Z}_k^{3/2}k\partial_k\bigl({{\partial U_k}\over
{\partial B^2}}\bigr)=-{g{\cal Z}_k^{3/2}\over 2B}{\partial\over{\partial B}}
\bigl(k\partial_k U_k\bigr) \cr
&
={g^3(k)\over {8\pi^2\tau}}\Biggl\{ {\rm ln}\bigl(1-\tau\bigr)+3{\rm ln}
\bigl(1+\tau\bigr)+4\sum_{n=1}^{\infty}{\rm ln}\bigl[1+(2n+1)\tau\bigr]
-2\sum_{n=0}^{\infty}{\rm ln}\bigl[1+(2n+1)\tau\bigr] \cr
&\quad
-\tau\Bigl[{1\over{1-\tau}}-{3\over{1+\tau}}-4\sum_{n=1}^{\infty}{(2n+1)\over
{1+(2n+1)\tau}}+2\sum_{n=0}^{\infty}{(2n+1)\over{1+(2n+1)\tau}}\Bigr]
\Biggr\},}}
where $\tau=gB/k^2$. It is rather interesting to note that the $\beta$
function here not only depends on $g(k)$, but also the dimensionless
parameter $\tau$. With the help of the Euler formula \nielsen:
\eqn\euler{ \sum_{n=0}^{\infty}h(n+{1\over 2})=\int_0^{\infty}dx~h(x)-{1\over
24}
h'(x)\Big\vert^{\infty}_0+\cdots, \quad\qquad h(\infty)=0,}
the $\beta$ function reads
\eqn\bbf{\beta(g(k),\tau)={g^3(k)\over 8\pi^2\tau}\Biggl\{
{\rm ln}\bigl(1-\tau\bigr)-{\rm ln}\bigl(1+\tau\bigr)+{\tau\over 3}
-{\tau\over 6}+\cdots
-\tau\Bigl[{1\over{1-\tau}}+{1\over{1+\tau}}-{1\over 3}
+{1\over 6}+\cdots\Bigr]\Biggr\},}
which in the limit of large $k$ or vanishing $\tau$, gives
\eqn\bbfu{ \beta(g(k))=-{g^3(k)\over 4\pi^2}\Bigl\{1+{2\over 3}
+{1\over 6}\Bigr\}=-{11g^3(k)\over 24\pi^2},}
in complete agreement with the that
obtained from \btafun\ for $SU(2)$. Notice that the contributions
to the $\beta$ function from the ``unstable mode'' (the first term)
and the ghost kernel (the third term) are, respectively, $-g^3(k)/4\pi^2$
and $-g^3(k)/24\pi^2$,
in accord with the analyses of Nielsen and Olsen \olsen.

We mentioned before that two multiplicity factors used in \maiani\ were
incorrect due to the negligence of the unphysical $S_z=0$
sector albeit the same $\beta$ function was given.
The way it was obtained is as follows: The original expression
which makes no reference of the unphysical sector $S_z=0$ actually gives
\eqn\by{ \beta_M(g(k))=-{g^3(k)\over 4\pi^2}\Bigl\{1+{3\over 4}
+{1\over 6}\Bigr\}=-{23g^3(k)\over 48\pi^2},}
instead of \bbfu. Removing the contribution from
the would-be unstable mode entirely by the subtraction
\eqn\remm{ \int{d^2p\over{(2\pi)^2}}
{\rm ln}\bigl(p^2+gB\bigr)-\int_{p^2>gB}{d^2p\over{(2\pi)^2}}{\rm ln}
\bigl(p^2-gB\bigr)}
followed by the substitution
\eqn\prol{ \int{ds\over s}={\rm ln}~s\longrightarrow 2\times{\rm ln}\bigl(
{\mu^2\over gB}\bigr)}
in the proper-time representation then gives the correct result:
\eqn\by{ \beta_M(g(k))\rightarrow -{g^3(k)\over 4\pi^2}~2\Bigl\{0+{3\over 4}
+{1\over 6}\Bigr\}=-{11g^3(k)\over 24\pi^2}.}
In other words, eliminating the contribution from the would-be
unstable mode completely followed by multiplying the extra factor of 2
as appeared on the right-hand-side of \prol\ is how the expected $\beta$
function was arrived in \maiani.

Taking into account the imaginary contribution which gets generated
for $p^2< gB$, the complete complex effective potential reads
\eqn\repot{ U={B^2\over 2}+{11g^2\over 48\pi^2}B^2\Bigl[
{\rm ln}\bigl({gB\over\mu^2}\bigr)-{1\over 2}\Bigr]-i{g^2B^2\over 8\pi^2},}
subject to the renormalization condition \matinyan
\eqn\renor{ {{\partial\bigl({\rm Re}~U\bigr)}\over{\partial\cal F}}\Big
\vert_{\tilde\mu^2}=1,}
with ${\cal F}=B^2/2$ being the gauge invariant quantity of the theory.
The condition is readily fulfilled by choosing $\tilde\mu^4=2g^2\cal F$.

The arbitrary scale $\mu^2$ and the IR cutoff $k$ arising from
the operator cutoff regularization can be related to each other by noting that
\eqn\repot{ \mu\partial_{\mu}\bigl({\rm Re}~U\bigr)
=-{11g^2B^2\over 24\pi^2},}
and
\eqn\loevy{k\partial_kU_k=-{k^2gB\over 4\pi^2}
\Bigl[{\rm ln}\Bigl({k^2-gB\over k^2+gB}\Bigr)
+2\sum_{n=0}^{\infty}{\rm ln}
\Bigl({{k^2+(2n+1)gB}\over k^2}\Bigr)\Bigr]\rightarrow
{11g^2B^2\over 24\pi^2}+{1\over 6\pi^2}{g^4B^4\over k^4}+\cdots}
in the large $k$ limit. This implies that we must have $\mu\partial_{\mu}
\equiv -k\partial_k$, i.e., the two scales run in the opposite manner.
This connection can also be seen from
\eqn\regor{ {1\over g^2(k)}={1\over g^2}-{11\over 24\pi^2}{\rm ln}
\Bigl({\tilde\Lambda^2\over k^2}\Bigr)}
given by \coupl\ and
\eqn\reggo{ {1\over g^2(\mu)}={1\over g^2}-{11\over 24\pi^2}{\rm ln}
\Bigl({\mu^2\over gB}\Bigr).}
In other words, with $k^2$ being the usual IR cutoff,
$\mu^2$ should be interpreted as an UV cutoff. By replacing
then right-hand-side of \loevy\ with the corresponding $k$-dependent running
parameters, the RG evolution equation for $U_k$ becomes
\eqn\runf{ k\partial_kU_k=-{2\beta(g(k))\over g(k)}U_k\Bigl(
1+{8g^2(k)\over 11}{U_k\over k^4}\Bigr)+\cdots.}
Solving this differential equation by retaining only the leading order
contribution, we have the following RG improved
blocked potential:
\eqn\soldf{ {\rm ln}U_k=-2\int{dk\over k}{\beta(g(k))\over g(k)}.}
Eq.\soldf\ is analogous to the result obtained in \matinyan\ by Matinyan and
Savvidy. However, it only takes into consideration the effect
the ${\cal F}=B^2/2$ term. In order to explore the influence of the higher
order operators, one must solve \runf\ completely without truncation.

We emphasize that the above treatments are limited to the regime where
$k$ is large and the theory is asymptotically free. Continuing to evolve
the system to a lower
$k$ will result in a complicated blocked action which invalidates
perturbation theory. In the IR region where $\tau$ is large,
eq.\runf\ is no longer a good approximation. Serious difficulties are
already encountered near $\tau= 1$ as the $\beta$ function in \bbt\ develops
pole and divergence. The source of the singularity is undoubtedly
due to the unstable mode which becomes unsuppressed for
$k \le \sqrt{gB}~(\tau \ge 1)$.
In \matinyan, Savvidy considered only the real part of
the potential in \repot\ and obtained a
nontrivial minimum:
\eqn\misol{ gB_{\min}=\mu^2e^{-24\pi^2/{11g^2}}.}
However, the existence of this vacuum configuration which lies in the
deep IR regime has been a subject
of intense debates for quite some time. The persistence of the unstable mode
in the IR region
lead Maiani {\it et al} to argue that the problem associated with
unstable configurations can only be treated nonperturbatively. On the other
hand, lattice calculations seem to support the formation of chromomagnetic
condensate \levi\ \ref\trottier. To provide a consistent check to
these claims, a successful nonperturbative RG approach
would be desirable. We therefore
propose to modify \evvy\ and write
\eqn\loevy{k\partial_kU_k=-{k^2g(k)\over 4\pi^2}\sqrt{2U_k}
\Biggl\{{\rm ln}\Bigl({{k^2-g(k)\sqrt{2U_k}}\over {k^2+g(k)\sqrt{2U_k}}}\Bigr)
+2\sum_{n=0}^{\infty}{\rm ln}
\Bigl({{k^2+(2n+1)g(k)\sqrt{2U_k}}\over k^2}\Bigr)\Biggr\},}
where the chromomagnetic field $B$ on the right-hand-side of \evvy\ has been
replaced by $Z_k^{1/2}\sqrt{2U_k}$ as suggested by \zze. Since no
assumption has been made on the value of $k$ in
\loevy, it remains valid even in the deep IR regime where $k$ is small.
Our nonlinear RG equation for the blocked potential $U_k$ provides
an improvement beyond the standard perturbative treatment
to the IR problem; in particular,
the RG dependence of the unstable mode can be explored.

\bigskip
\goodbreak
\medskip
\centerline{\bf VI. SUMMARY AND DISCUSSIONS}
\xdef\secsym{6.}\global\meqno = 1
\medskip

In this paper we derived the effective Yang-Mills blocked action
$\tilde S_k$ in a manifestly gauge invariant manner by the help
of the operator cutoff regularization.
The regulating smearing function ${\rho_k^{(d)}(s,\Lambda)}$ introduced
in the proper-time integration simulates a sharp momentum cutoff for
the leading order blocked potential $U_k$ and is
reminiscent to the Pauli-Villars regulator. The blocked action $\tilde S_k$
provides a smooth interpolation between the bare action defined at
$k=\Lambda$ and the effective action at an arbitrary energy scale $k$.
The presence of the IR scale $k$ readily allows us to study the RG
flow of the theory using the Wilson-Kadanoff blocking approach.

The conventional perturbative one-loop differential flow equation for the
Yang-Mills blocked action corresponds to the evolution for the gauge
coupling constant $g$. This is due to the fact that $g$ is the only
free parameter in the theory and the inclusion of the one-loop contributions
from both gauge and Faddeev-Popov ghost kernels is equivalent to
replacing $g$ by the running parameter $g(k)$. However, the situation changes
when we consider the RG improved equation \evy\ which
invariably
incorporates the contributions from other higher dimensional operators
that are being generated in the course of blocking transformation. These
operators, though initially absent in the bare lagrangian,
may play an important role in the effective lagrangian for the low
energy IR theory. The advantage of \evy\ is that it allows practical
calculations and it remains applicable even when $k$ is small.

In probing the vacuum structure of QCD, nonperturbative method
such as lattice gauge theory has often been employed. We find the RG
approach outlined in this paper a powerful alternative for investigating
the vacuum. Though the complicated nonlinear partial differential
equation written in \evy\ seems to make the analytical form for
the low energy blocked action rather hopeless, its numerical solutions may
nevertheless provide a consistent check for the lattice results.
For the simplest $SU(2)$ theory in the presence of a static chromomagnetic
field considered in Sec. V, a complete solution for the RG flow equation
\loevy\ may yield additional insights on the role of the unstable mode
as well as the effect of the higher order gauge
invariant operators such as ${\cal F}^2=B^4/4$. It may even
help resolve the longstanding issue of the reliability of the energetically
more favored ground state given in \misol.
For realistic theories, the
effects of matter fields too must be considered. Works along these directions
are now in progress.

\medskip
\goodbreak
\bigskip
\centerline{\bf ACKNOWLEDGEMENTS}
\medskip
\nobreak
We thank Professor J. Polonyi for initiating this project and continuous
guidance and encouragement. Valuable discussions with V. Branchina,
S. Matinyan and B. M\"uller
are greatly appreciated. This work is supported in part by funds
provided by the U. S. Department of Energy (D.O.E.) under contract
\#DE-FG05-90ER40592.

\medskip
\medskip
\goodbreak
\centerline{\bf APPENDIX A: SCALAR FIELD THEORY}
\xdef\secsym{\rm A.}\global\meqno = 1
\medskip

In this Appendix, we give the details of how blocked potentials for
scalar field theory are computed using operator cutoff formalism.
To be definite, the calculations will be carried out in $d=4$ dimension.
Consider for simplicity the following bare lagrangian:
\eqn\lagg{ {\cal L}={1\over 2}(\partial_{\mu}\phi)^2+V(\phi).}
In the presence of a slowly-varying background field $\Phi(x)$ whose
Fourier modes are constrained by an upper cutoff scale $k$, by
integrating out the fast-fluctuating modes, the one-loop contribution
to the low-energy blocked potential is given by
\eqn\htrc{\eqalign{U^{(1)}_k(\Phi)&=-{1\over 2}\int_0^\infty{ds\over s}
\rho_k^{(4)}(s,\Lambda)\int_p e^{-p^2s}\Bigl(e^{-V''(\Phi)s}-
e^{-V''(0)s}\Bigr)\cr
&=-{1\over 32\pi^2}\int_0^\infty{ds\over s^3}\rho_k^{(4)}(s,\Lambda)
\Bigl(e^{-V''(\Phi)s}-e^{-V''(0)s}\Bigr).}}
Notice that the scales set by momentum regularization
are now taken over by $\rho_k^{(4)}(s,\Lambda)$. As shown in \ocr, smearing
function of the form
\eqn\smres{\rho_k^{(4)}(s,\Lambda)=\Bigl[1-(1+\Lambda^2s)e^{-\Lambda^2s}\Bigr]
-\Bigl[1-(1+k^2s)e^{-k^2s}\Bigr]=\rho(\Lambda^2s)-\rho(k^2s)}
is equivalent to imposing sharp momentum cutoffs. That is, inserting \smres\
into \htrc\ leads to the cutoff expression:
\eqn\bpot{\eqalign{U^{(1)}_k(\Phi)&={1\over 2}\int_p^{'}{\rm ln}
\Bigl({{p^2+V''(\Phi)}\over {p^2+V''(0)}}\Bigr)={1\over64\pi^2}
\Biggl\{\bigl(\Lambda^2-k^2\bigr)\bigl(V''(\Phi)-V''(0)\bigr) \cr
&
+\Lambda^4{\rm ln}
\Bigl({{\Lambda^2+V''(\Phi)}\over {\Lambda^2+V''(0)}}\Bigr)
-k^4{\rm ln}\Bigl({{k^2+V''(\Phi)}\over {k^2+V''(0)}}\Bigr)
-V''(\Phi)^2{\rm ln}\Bigl({{\Lambda^2+V''(\Phi)}\over{k^2+V''(\Phi)}}
\Bigr)+ \cdots \Biggr\},}}
up to some $\Phi$-independent constant. Taking the $\lambda\phi^4$ theory
as an example, the blocked potential up to the one-loop order becomes
\eqn\oloop{\eqalign{U_k(\Phi) &=V(\Phi)-{1\over 2}\int_p
\int_0^\infty{ds\over s}\rho_k^{(4)}(s,\Lambda)e^{-(p^2+\mu_R^2)s}
\Bigl(e^{-\lambda_R\Phi^2s/2}-1\Bigr) \cr
&
={1\over 2}\Bigl[\mu^2+{\lambda_R\over 32\pi^2}\Bigl(\Lambda^2
+\mu_R^2{\rm ln}{\mu_R^2\over \Lambda^2}\Bigr)-{\lambda_R\over 64\pi^2}k^2
\Bigr]\Phi^2
+{1\over 4!}\Bigl[\lambda+{3\lambda_R^2\over 32\pi^2}
\Bigl(1+{\rm ln}{\mu_R^2\over\Lambda^2}\Bigr)\Bigr]\Phi^4 \cr
&
+{1\over 64\pi^2}\Bigl[\Bigl(\mu_R^2
+{\lambda_R\over 2}\Phi^2\Bigr)^2-k^4\Bigr]{\rm ln}\Bigl({{k^2+\mu_R^2
+\lambda_R\Phi^2/2}\over \mu_R^2}\Bigr) \cr
&
={\mu_R^2\over 2}\Phi^2\Bigl[1-{\lambda_R\over64\pi^2}
\Bigl(1+{k^2\over \mu_R^2}\Bigr)\Bigr]
+{\lambda_R\over 4!}\Phi^4\Bigl(1-{9\lambda_R\over64\pi^2}\Bigr)\cr
&
+{1\over64\pi^2}\Bigl[\Bigl(\mu_R^2+{\lambda_R\over2}\Phi^2\Bigr)^2-k^4\Bigr]
{\rm ln}\Bigl(1+{{k^2+\mu_R^2+\lambda_R\Phi^2/2}\over\mu_R^2}\Bigr),}}
where the renormalized parameters are given by
\eqn\wfrcs{\cases{\eqalign{ \mu_R^2&=\mu^2+{\lambda_R\over
32\pi^2}\Bigl(\Lambda^2+\mu_R^2{\rm ln}{\mu_R^2\over \Lambda^2}\Bigr) \cr
\lambda_R&=\lambda+{3\lambda_R^2\over 32\pi^2}
\Bigl(1+{\rm ln}{\mu_R^2\over\Lambda^2}\Bigr). \cr}}}
It is easily seen that in the limit $k=0$, \oloop\ reduces to the usual
effective potential obtained in \ref\coleman. For this theory, the
improved RG equation reads
\eqn\evmin{k\partial_kU_k(\Phi)=-{k^4\over16\pi^2}{\rm ln}\Bigl(
{{k^2+U''_k(\Phi)}\over {k^2+U''_k(0)}}\Bigr),}
which is a nonlinear differential equation that takes into account the
coupling between the high and the low momentum modes.

The results obtained above can be readily extended to scalar electrodynamics.
The lagrangian is given by
\eqn\qwlag{\eqalign{{\cal L}_{\rm SQED}=&-{1\over4}F_{\mu\nu}F_{\mu\nu}
-{1\over2\alpha}(\partial_\mu A_{\mu})^2\cr
&+\vert(\partial_\mu+ie_0A_\mu)\phi(x)\vert^2
+{\mu^2\over 2}\phi(x)^\dagger\phi(x)+{\lambda\over 6}
(\phi(x)^\dagger\phi(x))^2,\cr}}
where
$F_{\mu\nu}=\partial_\mu A_\nu-\partial_\nu A_\mu$ and $\alpha$ is the
gauge-fixing parameter.
The complex field $\phi(x)$ may be rewritten in terms of real fields
$\phi_1$ and $\phi_2$ as $(\phi_1(x)+i\phi_2(x))/{\sqrt 2}$.
Considering the special case where $A_c=0$ and $\Phi^a=\Phi\delta^{a,1}$
with $\Phi$ being the constant background configuration, the blocked
potential in the Landau gauge
with $\alpha=0$ becomes
\eqn\sqqe{\eqalign{ U_k^{(1)}(\Phi)&=
-{1\over 32\pi^2}\int_0^{\infty}{ds\over s^3}\Biggl\{e^{-\mu_R^2s}
\Bigl[(1+k^2s)e^{-k^2s}-(1+\Lambda^2s)e^{-\Lambda^2s}\Bigr] \cr
&
\times\bigl[\bigl(e^{-\lambda_R\Phi^2s/2}-1\bigr)
+\bigl(e^{-\lambda_R\Phi^2s/2}-1\bigr)\bigr]+3\bigl[1-(1+\Lambda^2s)
e^{-\Lambda^2s}\bigr]\bigl(e^{-e_0^2\Phi^2s}-1\bigr)\Biggr\}.}}
Even though blocking is performed only for the scalar fields,
it can be implemented in a similar fashion for gauge fields as well.
Notice that the extra factor of
three in the photon loop contribution arises
from the trace of the propagator in the Landau gauge. We also comment that
the form of $U_k(\Phi)$ is generally gauge dependent although
physical quantities must be gauge independent \ref\jackiw.

The theory, however, is plagued by IR singularity due to the
presence of massless photons. The problem could be avoided if blocking is also
done for the gauge fields, i.e., using $\rho_k^{(4)}(s,\Lambda)$
instead of $\rho_{k=0}^{(4)}(s,\Lambda)$. The conventional regularization
scheme is an off-shell subtraction condition for the coupling constant
\coleman:
\eqn\offs{ \lambda_R={{\partial^4 U_k(\Phi)}\over{\partial\Phi^4}}
\Big\vert_{\Phi=M,k=0}}
which, in the language of operator cutoff, is equivalent to using the
following coupling constant counterterm \ref\sergei:
\eqn\subl{\eqalign{\delta\lambda &={1\over 32\pi^2}
\int_0^{\infty}{ds\over s}\bigl[1-(1+\Lambda^2s)e^{-\Lambda^2s}\bigr]
\times\Biggl\{\lambda_R^2e^{-(\mu_R^2+\lambda_R M^2/2)s}\bigl(3-6\lambda_R M^2s
+\lambda_R^2M^4s^2\bigr) \cr
&\qquad\qquad
+{\lambda_R^2\over 81}e^{-(\mu_R^2+\lambda_R M^2/6)s}\bigl(27-18\lambda_R M^2s
+\lambda_R^2M^4s^2\bigr) \cr
&\qquad\qquad
+12e_0^4e^{-e_0^2M^2s}\bigl(3-12e_0^2M^2s+4e_0^4M^4s^2\bigr)\Biggr\}\cr
&
=-{1\over 64\pi^2}\Biggl\{{20\over 3}\lambda_R^2+{{4\lambda_R^3M^2(\lambda_R
M^2+9\mu_R^2)}\over{81(\mu_R^2+\lambda_RM^2/6)^2}}+{{4\lambda_R^3M^2(\lambda_R
M^2+3\mu_R^2)}\over{(\mu_R^2+\lambda_RM^2/2)^2}}\cr
&
+24e_0^4\Bigl[11+3{\rm ln}\bigl({M^2\over\Lambda^2}\bigr)\Bigr]+{2\lambda_R^2
\over 3}{\rm ln}\Bigl({{\mu_R^2+\lambda_RM^2/6}\over\Lambda^2}\Bigr)+6
\lambda_R^2{\rm ln}\Bigl({{\mu_R^2+\lambda_RM^2/2}\over\Lambda^2}\Bigr)
\Biggr\}.}}
After removing the $\Lambda$ dependence, the blocked potential becomes
\eqn\qeupk{\eqalign{U_k(\Phi)&= {\mu_R^2\over 2}\Phi^2+{\lambda_R\over 4!}
\Phi^4+{1\over 64\pi^2}\Biggl\{-{2\lambda_R\over 3}\bigl(k^2+\mu_R^2\bigr)
\Phi^2-{5\lambda_R^2\over 12}\Phi^4\cr
&
+\Bigl[\bigl(\mu_R^2+{\lambda_R\over 2}\Phi^2\bigr)^2-k^4\Bigr]{\rm ln}
\Bigl({{k^2+\mu_R^2+\lambda_R\Phi^2/2}\over\mu_R^2}\Bigr) \cr
&
+\Bigl[\bigl(\mu_R^2+{\lambda_R\over 6}\Phi^2\bigr)^2-k^4\Bigr]{\rm ln}
\Bigl({{k^2+\mu_R^2+\lambda_R\Phi^2/6}\over\mu_R^2}\Bigr)\cr
&
+{\lambda_R^2\over 4}\Phi^4~{\rm ln}\Bigl({\mu_R^2\over{\mu_R^2+\lambda_R
M^2/2}}\Bigr)+{\lambda_R^2\over 36}\Phi^4~{\rm ln}\Bigl({\mu_R^2\over
{\mu_R^2+\lambda_R M^2/6}}\Bigr)\cr
&
-{\lambda_R^3M^2\over 6}\Phi^4\biggl[{{\lambda_RM^2+9\mu_R^2}\over{81
\bigl(\mu_R^2 + \lambda_R M^2/6\bigr)^2}}+{{\lambda_RM^2+3\mu_R^2}\over
{\bigl(\mu_R^2+\lambda_R M^2/2\bigr)^2}}\biggr] \cr
&
+3e_0^4\Phi^4\Bigl[{\rm ln}\bigl({\Phi^2\over M^2}\bigr)-{25\over 6}\Bigr]
\Biggr\},}}
which for $\mu_R^2=k^2=0$ reduces to
\eqn\cde{U_{k=0}(\Phi)={\lambda_R\over 4!}\Phi^4+{\Phi^4\over 64\pi^2}
\bigl({5\over 18}\lambda_R^2+3e_0^4\bigr)\Bigl[{\rm ln}\bigl({\Phi^2\over
M^2}\bigr)-{25\over 6}\Bigr].}
The theory in this limit shows spontaneous symmetry
breaking driven by radiative corrections
\coleman. Once more, the symmetry-preserving nature of the
operator cutoff is seen from the absence of cutoff scales in
the $p$ integration.

\medskip
\bigskip
\centerline{\bf APPENDIX B: GENERALIZED PROPER-TIME CLASS}
\medskip
\nobreak
\xdef\secsym{{\rm B.}}\global\meqno = 1
\medskip
\nobreak

Numerous regularization schemes can all be shown to belong to the
generalized class of proper-time since they can be represented by
a suitable definition of
smearing function. A detailed discussion has already been given by Ball
\chiral\ and also in \zuk. However, for the sake of completeness
and comparative purpose,
we recapitulate here various examples
and examine how they modify the propagator and the
corresponding one-loop kernel. We also show how cutoff scales can be
implemented in dimensional regularization as well as $\zeta$ function
regularization. The ``hybrid'' prescriptions of dimensional cutoff
and $\zeta$ function cutoff allows us to establish a direct connection
with the momentum regularization. We specifically apply
each of these techniques to regularize the divergences encountered in
the computation of the two- and the four-point vertex functions for
${\cal H}=p^2+\mu^2$ in $d=4$.

\bigskip
\noindent{(1) operator cut-off:}

For $d=4$, the smearing function becomes
\eqn\urty{\eqalign{ \rho_k^{(4)}(s,\Lambda)&=(1+k^2s)e^{-k^2s}
-(1+\Lambda^2s)e^{-\Lambda^2s} \cr
&
=\bigl(e^{-k^2s}-e^{-\Lambda^2s}\bigr)+s\bigl(k^2e^{-k^2s}-\Lambda^2
e^{-\Lambda^2s}\bigr),}}
which leads to the following regularized propagator
and one-loop kernel:
\eqn\cher{\eqalign{{1\over {\cal H}^n}\Big\vert_{\rm oc}
&={1\over\Gamma(n)}\int_0^{\infty}ds~s^{n-1}\Bigl[(1+k^2s)e^{-k^2s}
-(1+\Lambda^2s)e^{-\Lambda^2s}\Bigr]e^{-{\cal H} s} \cr
&
={1\over{({\cal H}+k^2)}^n}-{1\over{({\cal H}+\Lambda^2)}^n}+{nk^2\over
{({\cal H}+k^2)^{n+1}}}-{n\Lambda^2\over{({\cal H}+\Lambda^2)^{n+1}}},}}
and
\eqn\hko{ {\rm Tr}_{\rm oc}{\rm ln}\Bigl( {{\cal H}\over {\cal H}_0}\Bigr)
={\rm Tr}\Biggl\{ {\rm ln}\Bigl[ { {{\cal H}+k^2}\over {{\cal H}_0+k^2}}\times
{{{\cal H}_0+\Lambda^2}\over {{\cal H}+\Lambda^2}}\Bigr]
-{\Lambda^2({\cal H}-{\cal H}_0)\over {({\cal H}+\Lambda^2)({\cal
H}_0+\Lambda^2)}}
+{k^2({\cal H}-{\cal H}_0)\over {({\cal H}+k^2)({\cal H}_0+k^2)}}\Biggr\} .}
As demonstrated in Sec. II, eq.\urty\ simulates a sharp cutoff at the
level of blocked potential. Using \htrc,
the one-loop correction to the two- and four-point vertex functions
for scalar theory can be written as
\eqn\sdfa{ \delta\Gamma^{(2)}_{\rm oc}={\partial^2U_k^{(1)}\over{
\partial\Phi^2}}
\Big\vert_{\Phi=0}={\lambda\over 32\pi^2}\int_0^{\infty}{ds\over s^2}
{\rho_k^{(4)}(s,\Lambda)} e^{-\mu^2s},}
and
\eqn\sdfu{ \delta\Gamma^{(4)}_{\rm oc}={\partial^4U_k^{(1)}\over{
\partial\Phi^4}}
\Big\vert_{\Phi=0}=-{3\lambda^2\over 32\pi^2}\int_0^{\infty}{ds\over s}
{\rho_k^{(4)}(s,\Lambda)} e^{-\mu^2s},}
which for $k=0$ become
\eqn\contr{\eqalign{\delta\Gamma^{(2)}_{\rm oc} &=\int_p
{1\over{p^2+\mu^2}}\Bigl({\Lambda^2\over{p^2+\mu^2+\Lambda^2}}\Bigr)^2 \cr
&
={\lambda\over 32\pi^2}
\int_0^{\infty}{ds\over s^2}\rho_{k=0}^{(4)}(s,\Lambda) e^{-\mu^2s}
={\lambda\over 32\pi^2}\Bigl[\Lambda^2-\mu^2{\rm ln}\Bigl({{\Lambda^2
+\mu^2}\over\mu^2}\Bigr)\Bigr],}}
and
\eqn\contry{\eqalign{\delta\Gamma^{(4)}_{\rm oc}&=
\int_p{1\over{(p^2+\mu^2)}^2}\Bigl({\Lambda^2\over{p^2+\mu^2+\Lambda^2}}
\Bigr)^2\Bigl[1+{2(p^2+\mu^2)\over{p^2+\mu^2+\Lambda^2}}\Bigr]\cr
&
=-{3\lambda\over 32\pi^2}
\int_0^{\infty}{ds\over s}\rho_{k=0}^{(4)}(s,\Lambda) e^{-\mu^2s}
=-{3\lambda^2\over 32\pi^2}\Bigl[{\rm ln}\Bigl({{\Lambda^2+\mu^2}
\over\mu^2}\Bigr)-{\Lambda^2\over{\mu^2+\Lambda^2}}\Bigr].}}
On the other hand, using the momentum cutoff procedure, one also has
\eqn\shcut{\delta\Gamma^{(2)}_{\Lambda}={\lambda\over 2}\int_p^{\Lambda}
{1\over{p^2+\mu^2}}
={\lambda\over 32\pi^2}\Bigl[\Lambda^2-\mu^2{\rm ln}\Bigl({{\Lambda^2
+\mu^2}\over\mu^2}\Bigr)\Bigr],}
and
\eqn\concu{\delta\Gamma^{(4)}_{\Lambda}=-{3\lambda^2\over 2}\int_p^{\Lambda}
{1\over{(p^2+\mu^2)}^2}=-{3\lambda^2\over 32\pi^2}\Bigl[{\rm ln}
\Bigl({{\Lambda^2+\mu^2}\over\mu^2}\Bigr)
-{\Lambda^2\over{\mu^2+\Lambda^2}}\Bigr].}

\bigskip
\noindent{(2) Pauli-Villars:}

The conventional Pauli-Villars scheme can be parameterized in the
proper-time representation by taking the smearing function to be
\eqn\pvsmear{ \rho_k^{\rm pv}(s,\Lambda)=\sum_i\Bigl(a_ie^{-k_i^2s}
-b_ie^{-\Lambda_i^2s}\Bigr),}
where $\Lambda_i$ are the masses of some ghost
states, and $k_i$ the extra masses added to the spectra.
To render the theory finite, the
coefficients $a_i$ and $b_i$ as well as $i$, the number
of ghost terms are appropriately chosen.
Physical limit, however, corresponds to
taking $\Lambda_i\to\infty$ and $k_i\to 0$ since $\Lambda_i$ and $k_i$
control, respectively, the divergent behaviors of the theory in the UV and
the IR regimes. Eq. \pvsmear\ implies
\eqn\chpv{{1\over {\cal H}^n}\Big\vert_{\rm pv}=\sum_i{1\over\Gamma(n)}
\int_0^{\infty}ds~s^{n-1}\Bigl(a_ie^{-k_i^2s}-b_ie^{-\Lambda_i^2s}\Bigr)
e^{-{\cal H} s}=\sum_i\Bigl[{a_i\over{({\cal H}+k_i^2)^n}}-{b_i\over{({\cal H}
+\Lambda_i^2)^n}}\Bigr],}
and
\eqn\hpv{\eqalign{{\rm Tr}_{\rm pv}{\rm ln}\Bigl({{\cal H}\over {\cal H}_0}
\Bigr)&=-\sum_i\int_0^{\infty}{ds\over s}\Bigl(a_ie^{-k_i^2s}
-b_ie^{-\Lambda_i^2s}\Bigr){\rm Tr}\Bigl(e^{-{\cal H} s}-e^{-{\cal H}_0
s}\Bigr) \cr
&
={\rm Tr}\sum_i{\rm ln}\Biggl[\Bigl({{{\cal H}+k_i^2}\over{{\cal H}_0
+k_i^2}}\Bigr)^{a_i}\times \Bigl(
{{{\cal H}_0+\Lambda_i^2}\over {{\cal H}+\Lambda_i^2}}\Bigr)^{b_i}\Biggr].}}
The similarity between the operator cutoff and the Pauli-Villars is now
apparent. By choosing $a_i=b_i=i=1$, we notice
that the two smearing functions differ from one another only by a higher
order correction.

In computing $\delta\Gamma^{(2)}_{\rm pv}$ using the
Pauli-Villars regulator, it is necessary to introduce two ghost terms
since the integral in \shcut\ is quadratically divergent.
Thus, we write \ref\cheng
\eqn\pvtwo{\eqalign{ {1\over{p^2+\mu^2}}\Big\vert_{\rm pv}&=
{1\over{p^2+\mu^2}}-{b_1\over{p^2+\mu^2+\Lambda_1^2}}-{b_2\over{p^2+\mu^2
+\Lambda_2^2}} \cr
&
={f(p^2,\mu^2,\Lambda_1^2,\Lambda_2^2)\over{(p^2+\mu^2)(p^2+\mu^2+\Lambda_1^2)
(p^2+\mu^2+\Lambda_2^2)}},}}
where
\eqn\numera{\eqalign{f(p^2,\mu^2,\Lambda_1^2,\Lambda_2^2)&=
\bigl(1-b_1-b_2\bigr)p^4+\Bigl[2\bigl(1-b_1-b_2\bigr)\mu^2+\bigl(1
-b_1\bigr)\Lambda_2^2+\bigl(1-b_2\bigr)\Lambda_1^2\Bigr]p^2\cr
&
+\mu^2\Bigl[\bigl(1-b_1\bigr)\Lambda_2^2+\bigl(1-b_2\bigr)\Lambda_1^2\Bigr]
+\Lambda_1^2\Lambda_2^2,}}
and demand that
\eqn\edfr{ {1\over{p^2+\mu^2}}\Big\vert_{\rm pv}\longrightarrow {1\over p^6},
\qquad\qquad {\rm as}~~{p^2\to\infty}.}
The condition is satisfied if
\eqn\sewrt{ b_1+b_2-1=0, \qquad\qquad (1-b_2)\Lambda_1^2
+(1-b_1)\Lambda_2^2=0,}
which implies
\eqn\choer{ b_1={\Lambda_2^2\over{\Lambda_2^2-\Lambda_1^2}},\qquad\qquad
b_2=-{\Lambda_1^2\over{\Lambda_2^2-\Lambda_1^2}}.}
The correction to the two-point function can now be obtained as
\eqn\twop{\eqalign{ \delta\Gamma_{\rm pv}^{(2)}&=\int_p~{1\over{p^2+\mu^2}}
\Big\vert_{\rm pv}=\int_p~{\Lambda_1^2\Lambda_2^2\over
{(p^2+\mu^2)(p^2+\mu^2+\Lambda_1^2)(p^2+\mu^2+\Lambda_2^2)}} \cr
&
\longrightarrow~\int_p
{1\over{p^2+\mu^2}}\Bigl({\Lambda^2\over{p^2+\mu^2+\Lambda^2}}\Bigr)^2
={1\over 16\pi^2}\Bigl[\Lambda^2-\mu^2{\rm ln}\Bigl({{\Lambda^2
+\mu^2}\over\mu^2}\Bigr)\Bigr],~~~{(\Lambda_1,\Lambda_2\to\Lambda)},}}
which is in agreement with \shcut. As for
$\delta\Gamma_{\rm pv}^{(4)}$, since it is logarithmically divergent,
only one ghost term is sufficient and we obtain
\eqn\fourpv{\eqalign{ \delta\Gamma_{\rm pv}^{(4)}&=\int_p~{1\over{(p^2
+\mu^2)^2}}\Big\vert_{\rm pv}=\int_p\Bigl[{1\over{(p^2+\mu^2)^2}}-
{1\over{(p^2+\mu^2+\Lambda^2)^2}}\Bigr]
={1\over 16\pi^2}~{\rm ln}\Bigl({{\Lambda^2+\mu^2}\over\mu^2}\Bigr).}}

\bigskip
\noindent{(3) proper-time cutoff:}

Since divergences generated from taking
the trace in space-time are transferred into singularities in the proper-time
integration, one may regularize the theory by a direct truncation of the
integration regime(s) to avoid singularity. For example, we
may simply take the smearing function to be a sharp proper-time cutoff:
\eqn\psmer{ \rho_k^{\rm pc}(s,\Lambda)=\Theta(s-{1\over\Lambda^2})\Theta
({1\over k^2}-s).}
In this manner, we have
\eqn\hregg{\eqalign{ {1\over {\cal H}^n}\Big\vert_{\rm pc}
&={1\over\Gamma(n)}\int_0^{\infty}ds~s^{n-1}\Theta(s-{1\over\Lambda^2})
\Theta({1\over k^2}-s)e^{-{\cal H} s}
={1\over\Gamma(n)}\int_{1/{\Lambda^2}}^{1/k^2}ds~s^{n-1}e^{-{\cal H} s} \cr
&
={1\over{\cal H}^n}\cdot{1\over\Gamma(n)}\Bigl(
\Gamma\bigl[n,0,{{\cal H}\over k^2}\bigr]-\Gamma\bigl[n,0,{{\cal
H}\over\Lambda^2}
\bigr]\Bigr),}}
and
\eqn\hpve{\eqalign{{\rm Tr}_{\rm pc}{\rm ln}\Bigl({{\cal H}\over
{\cal H}_0}\Bigr)&=-\int_{1/{\Lambda^2}}^{1/k^2}{ds\over s}
{\rm Tr}\Bigl(e^{-{\cal H} s}-e^{-{\cal H}_0 s}\Bigr) \cr
&
={\rm Tr}\Biggl\{-{\rm Ei}({-{\cal H}/k^2})+{\rm Ei}({-{\cal H}_0/k^2})
+{\rm Ei}({-{\cal H}/{\Lambda^2}})-{\rm Ei}({-{\cal H}_0/{\Lambda^2}})\Biggr\}
\cr
&
={\rm Tr}{\rm ln}\Bigl({{\cal H}\over {\cal H}_0}\Bigr)+\cdots ,}}
where we have employed the asymptotic forms of the
exponential-integral function
\eqn\ertf{ {\rm Ei}(-s_0)=-\int_{s_0}^{\infty}{ds\over s}e^{-s}
=\cases{ {\rm ln}{s_0}+\gamma+\sum_{n=1}^{\infty}{{(-s_0)^n}\over n!n}
&$( s_0\to 0^{+})$, \cr
\cr
-{e^{-s_0}\over s_0} &$(s_0\to\infty)$.\cr}}
Correspondingly, we have
\eqn\selff{\delta\Gamma^{(2)}_{\rm pc}=\int_p{e^{-(p^2+\mu^2)/
{\Lambda^2}}\over{p^2+\mu^2}}={1\over 16\pi^2}\Bigl[\Lambda^2
e^{-\mu^2/{\Lambda^2}}+\mu^2{\rm Ei}(-{\mu^2\over\Lambda^2})\Bigr]
={1\over 16\pi^2}\Bigl[\Lambda^2-\mu^2{\rm ln}\Bigl({\Lambda^2\over
\mu^2}\Bigr)\Bigr]+\cdots,}
and
\eqn\selw{\eqalign{\delta\Gamma^{(4)}_{\rm pc}&=\int_p
{e^{-(p^2+\mu^2)/{\Lambda^2}}\over{(p^2+\mu^2)}^2}\Bigl[1+{{p^2+\mu^2}
\over\Lambda^2}\Bigr] \cr
&
={1\over 16\pi^2}\Bigl[-\bigl(1-{2\mu^2\over\Lambda^2}-{2\mu^4\over\Lambda^4}
\bigr){\rm Ei}(-{\mu^2\over\Lambda^2})+2(1+{\mu^2\over\Lambda^2})
e^{-\mu^2/{\Lambda^2}}\Bigr]
={1\over 16\pi^2}{\rm ln}\Bigl({\Lambda^2\over\mu^2}\Bigr)+\cdots.}}

\bigskip
\noindent{(4) point-splitting:}

One may also choose the smearing function to be of the form
\eqn\smeew{ \rho_k^{\rm ps}(s,\Lambda)=e^{-1/{\Lambda^2 s}}-e^{-1/{k^2 s}},}
which corresponds to the so-called point-splitting regularization scheme.
This smearing function yields
\eqn\serd{\eqalign{ {1\over {\cal H}^n}\Big\vert_{\rm ps}&={1\over\Gamma(n)}
\int_0^{\infty}ds~s^{n-1}\bigl(e^{-1/{\Lambda^2 s}}-e^{-1/{k^2 s}}\bigr)
e^{-{\cal H} s} \cr
&
={1\over{\cal H}^n}\cdot{2\over\Gamma(n)}\Bigl[\Bigl({{\cal H}\over\Lambda^2}
\Bigr)^{n/2}K_{n}\bigl({2{\cal H}^{1/2}\over\Lambda}\bigr)
-\Bigl({{\cal H}\over k^2}\Bigr)^{n/2}K_{n}\bigl({2{\cal H}^{1/2}\over
k}\bigr)\Bigr]
={1\over{\cal H}^n}+\cdots,}}
and
\eqn\hpve{\eqalign{{\rm Tr}_{\rm ps}{\rm ln}\Bigl({{\cal H}\over
{\cal H}_0}\Bigr)&=-\int_0^{\infty}{ds\over s}\bigl(
e^{-1/{\Lambda^2 s}}-e^{-1/{k^2 s}}\bigr){\rm Tr}\Bigl(e^{-{\cal H} s}
-e^{-{\cal H}_0 s}\Bigr) \cr
&
=2{\rm Tr}\Bigl[K_0\bigl({2{\cal H}_0^{1/2}\over\Lambda}\bigr)
-K_0\bigl({2{\cal H}^{1/2}
\over\Lambda}\bigr)-K_0\bigl({2{\cal H}_0^{1/2}\over k}\bigr)
+K_0\bigl({2{\cal H}^{1/2}\over k}\bigr)\Bigr] \cr
&
={\rm Tr}{\rm ln}\Bigl({{\cal H}\over {\cal H}_0}\Bigr)+\cdots,}}
where we have expanded the modified Bessel function asymptotically
as \ref\arfken:
\eqn\knxz{K_n(x)\sim \sqrt{\pi\over 2x}e^{-x}\Bigl[1+{(4n^2-1^2)\over{1!8x}}
+{{(4n^2-1^2)(4n^2-3^2)}\over{2!(8x)^2}}+\cdots\Bigr]\qquad (x\to\infty),}
and
\eqn\knxzw{K_n(x)\sim\cases{ 2^{n-1}(n-1)!x^{-n}+\cdots &$n\ge 1$,\cr
\cr
-{\rm ln}{x\over 2}-\gamma &$n=0$.\cr}\qquad\qquad (x\to 0^{+}). }
The two- and four-point functions in this scheme are as follows:
\eqn\twops{\eqalign{\delta\Gamma^{(2)}_{\rm ps}&=\int_p{1\over{p^2+\mu^2}}
\Big\vert_{\rm ps}=\int_0^{\infty}dy~e^{-y-\mu^2/{\Lambda^2y}}
\int_p{e^{-p^2/{\Lambda^2y}}\over{p^2+\mu^2}}\cr
&
={1\over 16\pi^2}\int_0^{\infty}dy~e^{-y-\mu^2/{\Lambda^2y}}\Bigl[\Lambda^2y
+\mu^2e^{\mu^2/{\Lambda^2y}}~{\rm Ei}\bigl(-{\mu^2\over{\Lambda^2y}}\bigr)
\Bigr] \cr
&
\approx{\mu^2\over 16\pi^2}\Bigl\{2K_2\bigl({2\mu\over\Lambda}\bigr)
+\int_0^{\infty}dy
e^{-y}~{\rm ln}\Bigl({\mu^2\over{\Lambda^2y}}\Bigr)\Bigr\}
={1\over 16\pi^2}\Bigl[\Lambda^2-\mu^2{\rm ln}\Bigl({{\Lambda^2
+\mu^2}\over\mu^2}\Bigr)\Bigr]+\cdots,}}
and
\eqn\selw{\eqalign{\delta\Gamma^{(4)}_{\rm ps}&=\int_p{1\over{(p^2+\mu^2)^2}}
\Big\vert_{\rm ps}=\int_0^{\infty}dy~ye^{-y-\mu^2/{\Lambda^2y}}
\int_p{e^{-p^2/{\Lambda^2y}}\over{(p^2+\mu^2)^2}} \cr
&
=-{1\over 16\pi^2}\int_0^{\infty}dy~ye^{-y-\mu^2/{\Lambda^2y}}\Bigl\{
1+\bigl(1+{\mu^2\over{\Lambda^2y}}\bigr)e^{\mu^2/{\Lambda^2y}}
{}~{\rm Ei}\bigl(-{\mu^2\over{\Lambda^2y}}\bigr)\Bigr\} \cr
&
={1\over 16\pi^2}{\rm ln}
\Bigl({{\Lambda^2+\mu^2}\over\mu^2}\Bigr)+\cdots.}}

We remark that the four smearing functions presented so far in a certain
sense can all be
viewed as a special case of the generalized momentum regularization in
which the
regularized integral for an arbitrary momentum-dependent function $f(p)$
is written as
\eqn\regfd{ \int_{b(k)}^{a(\Lambda)}dp f(p),}
where $a(\Lambda)$ and $b(k)$ are arbitrary functions of the cutoffs $\Lambda$
and $k$, respectively. This is readily seen by noting that the prescriptions
presented previously can be related to the generalized momentum
regularization via
\eqn\pref{ \rho_k^{\rm reg.}(s,\Lambda)\int_pe^{-p^2s}={1\over({4\pi s})^{d/2}}
\rho_k^{\rm reg.}(s,\Lambda)=\int_p^{'}e^{-p^2s}=S_d\int_{b(k)}^{a(\Lambda)}
dp~p^{d-1}e^{-p^2s},}
or
\eqn\regt{ \rho_k^{\rm reg.}(s,\Lambda)={2s^{d/2}\over\Gamma(d/2)}
\int_{b(k)}^{a(\Lambda)}dp~p^{d-1}e^{-p^2s}.}
For example, in the $d=4$ Pauli-Villars case, we have
\eqn\rupv{\eqalign{ \rho_k^{\rm pv}(s,\Lambda)&=e^{-k^2s}-e^{-\Lambda^2s}
=2s^2\int_{b(k)}^{a(\Lambda)}dp~p^3e^{-p^2s} \cr
&
=\bigl(1+b^2(k)s\bigr)e^{-b^2(k)s}
-\bigl(1+a^2(\Lambda)s\bigr)e^{-a^2(\Lambda)s},}}
where $a(\Lambda)$ obeys the transcendental equation
\eqn\afds{ e^{-\Lambda^2s}=\bigl(1+a^2(\Lambda)s\bigr)e^{-a^2(\Lambda)s}.}
The IR cutoff function $b(k)$ can be obtained in a similar manner.

\bigskip
\noindent{(5.a) dimensional regularization:}

One can also show that dimensional regularization falls into the generalized
class of proper-time by taking the smearing function to be
\eqn\dimr{\rho_{\epsilon}(s)=\bigl({4\pi s}\bigr)^{\epsilon/2},}
which suggests
\eqn\serw{ {1\over {\cal H}^n}\Big\vert_{\epsilon}={(4\pi)^{\epsilon/2}\over
\Gamma(n)}\int_0^{\infty}ds~s^{\epsilon/2+n-1}e^{-{\cal H} s}
={\Gamma(n+{\epsilon/2})\over\Gamma(n)}(4\pi)^{\epsilon/2}
{\cal H}^{-(n+\epsilon/2)},}
and
\eqn\hpve{\eqalign{{\rm Tr}_{\epsilon}{\rm ln}\Bigl({{\cal H}\over
{\cal H}_0}\Bigr)&=-(4\pi)^{\epsilon/2}\int_0^{\infty}ds
s^{-1+\epsilon/2}{\rm Tr}\Bigl(e^{-{\cal H} s}-e^{-{\cal H}_0 s}\Bigr) \cr
&
=-(4\pi)^{\epsilon/2}\Gamma({\epsilon/2}){\rm Tr}\Bigl({\cal H}^{-\epsilon/2}
-{\cal H}_0^{-\epsilon/2}\Bigr).}}
We remark here that this proper-time version of dimensional regularization
differs from the conventional one in the sense that calculations
are done in the original dimension. The manner in which the theory is
regularized is to increase the power of the propagator ${\cal H}^{-1}$
by $\epsilon/2$, thereby decreasing the degree of divergence.
The advantage of using this method is that no difficulty is encountered
when dealing with objects such as $\gamma_5$ which are defined on a
specific dimension. In fact, this corresponds to adding extra degrees
of freedom that are not directly coupled to the background fields \chiral.

The corrections to the two- and four-point functions are, respectively,
\eqn\twodim{\eqalign{\delta\Gamma^{(2)}_{\epsilon}&=\int_p{1\over{p^2+\mu^2}}
\Big\vert_{\epsilon}=(4\pi)^{\epsilon/2}\int_0^{\infty}ds~s^{\epsilon/2}
e^{-\mu^2s}\int_p e^{-p^2s} \cr
&
={1\over{(4\pi)^{2-\epsilon/2}}}\bigl(\mu^2\bigr)^{1-\epsilon/2}
\Gamma(-1+\epsilon/2)
={1\over 16\pi^2}\Bigl[-{2\mu^2\over\epsilon}-\mu^2{\rm ln}\Bigl({4\pi\over
\mu^2}\Bigr)\Bigr]+\cdots,}}
and
\eqn\fdim{\delta\Gamma^{(4)}_{\epsilon}=\int_p{1\over{(p^2+\mu^2)^2}}
\Big\vert_{\epsilon}={1\over{(4\pi)^{2-\epsilon/2}}}\bigl(
\mu^2\bigr)^{-\epsilon/2}\Gamma(\epsilon/2)
={1\over 16\pi^2}\Bigl[{2\over\epsilon}+{\rm ln}\Bigl({4\pi\over\mu^2}\Bigr)
\Bigr]+\cdots,}
where we have used \ref\ryder
\eqn\dimee{ \Gamma(-n+\epsilon/2)={(-1)^n\over n!}\Bigl[{2\over\epsilon}
+\bigl(1+{1\over 2}+\cdots+{1\over n}-\gamma\bigr)+O(\epsilon)\Bigr].}
The divergences now appear as poles for $\epsilon=0$ and $2$ since
\eqn\ferte{ \Gamma(-1+\epsilon/2)={1\over{\bigl(-1+\epsilon/2\bigr)
\bigl(\epsilon/2\bigr)}}\Gamma(1+\epsilon/2).}
These poles can be mapped onto the divergent expressions obtained using
the momentum cutoffs.

\bigskip
\noindent{(5.b) dimensional cutoff regularization:}

The most direct way to establish the connection between dimensional
regularization and the momentum cutoff regulator is by means of
the ``dimensional cutoff regularization'' defined by
\eqn\dimru{\rho_{\epsilon'}(s,\Lambda)=
\rho_{\epsilon}(s)\rho_k^{(d)}(s,\Lambda)=
{({4\pi})^{\epsilon/2}\over{S_d\Gamma(d/2)}}s^{(d+\epsilon)/2}\int_z^{'}
e^{-z^2s},}
which is simply the product of the two
smearing functions taken from each scheme
in $d$ dimension. The modified propagator and kernel in this $\epsilon'$ scheme
read
\eqn\hkkw{\eqalign{ {1\over {\cal H}^n}\Big\vert_{\epsilon'}&
={1\over\Gamma(n)}
\int_0^\infty ds~s^{n-1}\rho_k^{(d)}(s,\Lambda)e^{-{\cal H}s} \cr
&
={1\over{\cal H}^n}\cdot{{2(4\pi)^{\epsilon/2}\Gamma(n+d/2)}\over
{d\Gamma(n)\Gamma(d/2)}}
\Biggl\{\Bigl({\Lambda^2\over{\cal H}}\Bigr)^{d/2}F\Bigl({d\over 2},
{{d+\epsilon}\over 2}+n,1+{d\over 2};-{\Lambda^2\over{\cal H}}\Bigr) \cr
&\qquad\qquad\qquad\qquad\qquad~~
-\Bigl({k^2\over{\cal H}}\Bigr)^{d/2}F\Bigl({d\over 2},
{{d+\epsilon}\over 2}+n,1+{d\over 2};-{k^2\over{\cal H}}\Bigr)\Biggr\},}}
and
\eqn\hkosr{\eqalign{ &{\rm Tr}_{\epsilon'}{\rm ln}\Bigl(
{{\cal H}\over {\cal H}_0}\Bigr)=-\int_0^\infty{ds\over
s}{\rho_k^{(d)}(s,\Lambda)}
{\rm Tr}\Bigl(e^{-{\cal H}s}-e^{-{\cal H}_0s}\Bigr) \cr
&~
=-{2(4\pi)^{\epsilon/2}\over d}{\rm Tr}\Bigl\{\bigl({\Lambda^2\over{\cal H}}
\bigr)^{d/2}F\bigl({d\over 2},
{{d+\epsilon}\over 2},1+{d\over 2};-{\Lambda^2\over{\cal H}}\bigr)
-\bigl({\Lambda^2\over{\cal H}_0}\bigr)^{d/2}F\bigl({d\over 2},
{{d+\epsilon}\over 2},1+{d\over 2};-{\Lambda^2\over{\cal H}_0}\bigr) \cr
&\qquad\qquad
-\bigl({k^2\over{\cal H}}\bigr)^{d/2}F\bigl({d\over 2},
{{d+\epsilon}\over 2},1+{d\over 2};-{k^2\over{\cal H}}\bigr)
+\bigl({k^2\over{\cal H}_0}\bigr)^{d/2}F\bigl({d\over 2},
{{d+\epsilon}\over 2},1+{d\over 2};-{k^2\over{\cal H}_0}\bigr)\Bigr\}.}}
Eq. \twodim\ and \fdim\ are now modified as:
\eqn\twomk{\eqalign{\delta\Gamma^{(2)}_{\epsilon'}&
=\int_p{1\over{p^2+\mu^2}}
\Big\vert_{\epsilon'}=(4\pi)^{\epsilon/2}\int_0^{\infty}
ds~s^{\epsilon/2}
\bigl[1-(1+\Lambda^2s)e^{-\Lambda^2s}\bigr]e^{-\mu^2s}\int_p e^{-p^2s} \cr
&
={\Gamma(-1+\epsilon/2)\over{(4\pi)^{2-\epsilon/2}}}
\Biggl\{\bigl(\mu^2\bigr)^{1-\epsilon/2}-{{\epsilon\Lambda^2/2+\mu^2}\over
{(\Lambda^2+\mu^2)^{\epsilon/2}}}\Biggr\} ,}}
and
\eqn\fdimy{\delta\Gamma^{(4)}_{\epsilon'}=\int_p{1\over{
(p^2+\mu^2)^2}}\Big\vert_{\epsilon'}={\Gamma(\epsilon/2)
\over{(4\pi)^{2-\epsilon/2}}}
\Biggl\{\bigl(\mu^2\bigr)^{-\epsilon/2}-{{(1+\epsilon/2)\Lambda^2+\mu^2}\over
{(\Lambda^2+\mu^2)^{1+\epsilon/2}}}\Biggr\}.}
To recover the cutoff results, we take the limit $\epsilon\to 0$ first and
obtain
\eqn\twui{\eqalign{\delta\Gamma^{(2)}_{\epsilon'}&=-{\mu^2\over 8\pi^2}
\Biggl\{\Bigl[{1\over\epsilon}-{1\over 2}{\rm ln}{4\pi\mu^2}
\Bigr]-\Bigl[{1\over\epsilon}+{\Lambda^2\over 2\mu^2}-{1\over 2}{\rm ln}
{4\pi(\Lambda^2+\mu^2)}\Bigr]\Biggr\}+O(\epsilon) \cr
&
={1\over 16\pi^2}\Bigl[\Lambda^2-\mu^2{\rm ln}\Bigl({{\Lambda^2
+\mu^2}\over\mu^2}\Bigr)\Bigr]+O(\epsilon),}}
and
\eqn\fdimw{\eqalign{\delta\Gamma^{(4)}_{\epsilon'}&={1\over
16\pi^2}\bigl(
{2\over\epsilon}-1+{\rm ln}{4\pi}\bigr)\Biggl\{\Bigl[1-{\epsilon\over 2}
{\rm ln}\mu^2\Bigr]-\Bigl[1-{\epsilon\over 2}{\rm ln}\bigl(\Lambda^2
+\mu^2\bigr)+{{\epsilon\Lambda^2/2}\over{\Lambda^2+\mu^2}}\Bigr]\Biggr\}
+\cdots \cr
&
={1\over 16\pi^2}\Bigl[{\rm ln}\Bigl({{\Lambda^2+\mu^2}\over\mu^2}\Bigr)
-{\Lambda^2\over{\mu^2+\Lambda^2}}\Bigr]+O(\epsilon).}}
The above equations explicitly demonstrate how with this hybrid
dimensional cutoff regulator,  the $1/{\epsilon}$ singular
term coming from dimensional regularization and cutoff regularization
cancels each other and gives back the $\Lambda$ dependence of the cutoff
theory shown in \shcut\ and \concu. On the other hand, taking the limit
$\Lambda\to\infty$ before $\epsilon\to 0$ allows us to recover the usual
dimensional regularization scheme. In other words, depending on the order
in which the limits $\Lambda\to\infty$ and $\epsilon\to 0$ are taken,
different regularization schemes are actually achieved.

\bigskip
\noindent{(6.a) $\zeta$-function regularization:}

$\zeta$ function regularization has been discussed extensively by
Elizalde {\it et al}
\ref\elizalde\ and in the
context of operator regularization by McKeon {\it et al}
\mckeon.

In the $\zeta$-function regularization, the logarithm of an operator is
represented by
\eqn\sdf{ {\rm ln}~{\cal H}=-\lim_{t\to 0}{d\over dt}~{\cal H}^{-t}.}
Noting that
\eqn\ssfd{{1\over {\cal H}^{t}}={1\over\Gamma(t)}\int_0^{\infty}ds~s^{t-1}
e^{-{\cal H} s},}
one may define the $\zeta$-function as
\eqn\zzt{ \zeta(t)={1\over\Gamma(t)}\int_0^{\infty}ds~s^{t-1}~
{\rm Tr}~e^{-{\cal H} s},}
which implies
\eqn\ddsdf{ {\rm det}~{\cal H} ={\rm exp}\Bigl[{\rm Tr~ln}~{\cal H}\Bigr]
={\rm exp}\Biggl\{ {\rm Tr}~\lim_{t\to 0}\Bigl[-{d\over dt}~{\cal H}^{-t}
\Bigr]\Biggr\}
={\rm exp}\Bigl[-\lim_{t\to 0}{d\over dt}\zeta(t)\Bigr].}

The equivalent of $\zeta$-function regularization in the proper-time
formulation can be obtained by choosing the following smearing function:
\eqn\proo{ \rho_t^{\zeta}(s)=\lim_{t\to 0}{d\over dt}{1\over\Gamma(t)}
s^t,}
which gives
\eqn\serw{\eqalign{ {1\over {\cal H}^n}\Big\vert_{\zeta}&=\lim_{t\to 0}
{d\over dt}\Biggl\{
{1\over\Gamma(t)\Gamma(n)}\int_0^{\infty}ds~s^{n+t-1}e^{-{\cal H} s}\Biggr\}
=\lim_{t\to 0}{d\over dt}\Biggl\{ {\Gamma(n+t)\over{\Gamma(n)\Gamma(t)}}
{\cal H}^{-(n+t)}\Biggr\} \cr
&
=\lim_{t\to 0}\Biggl\{{\Gamma(n+t)\over{\Gamma(n)\Gamma(t)}}{\cal H}^{-(n+t)}
\Bigl[\psi(n+t)-\psi(t)-{\rm ln}{\cal H}\Bigr]\Biggr\}\cr
&
={1\over {\cal H}^n}
\cdot\lim_{t\to 0}\Biggl\{{\Gamma(n+t)\over{\Gamma(n)\Gamma(t+1)}}
{\cal H}^{-t}\Bigl[1+t\Bigl(\sum_{\ell=1}^{n-1}{1\over{t+\ell}}-{\rm ln}{\cal
H}
\Bigr)\Bigr]\Biggr\}
\longrightarrow {1\over {\cal H}^n},}}
and
\eqn\hpw{\eqalign{{\rm Tr}_{\zeta}{\rm ln}\Bigl({{\cal H}\over
{\cal H}_0}\Bigr)&=-\lim_{t\to 0}{d\over dt}\Biggl\{{1\over\Gamma(t)}
\int_0^{\infty}ds~s^{t-1}{\rm Tr}\Bigl(e^{-{\cal H} s}-e^{-{\cal H}_0 s}\Bigr)
\Biggr\} \cr
&
={\rm Tr}\biggl[-\lim_{t\to 0}{d\over dt}\Bigl({\cal H}^{-t}-{\cal
H}_0^{-t}\Bigr)
\biggr]={\rm Tr}{\rm ln}\Bigl({{\cal H}\over{\cal H}_0}\Bigr),}}
where we have used \ref\gared
\eqn\cder{\psi(x)={\Gamma'(x)\over\Gamma(x)},}
and
\eqn\fertw{ \psi(n+t)=\psi(t)+\sum_{\ell=0}^{n-1}{1\over{t+\ell}}.}
\bigskip
\noindent{(6.b) $\zeta$-function cutoff regularization:}

In an analogous manner to the dimensional cutoff regularization scheme, one
can introduce cutoff scales to the
$\zeta$-function cutoff regularization as well.
This again can be done with the product of two smearing functions:
\eqn\prg{ \rho_{t,k}^{\zeta,(d)}(s,\Lambda)=\rho_t^{\zeta}(s)\rho_k^{(d)}(s,
\Lambda)=\lim_{t\to 0}{d\over
dt}{1\over\Gamma(t)}s^t{\rho_k^{(d)}(s,\Lambda)}.}

To show that the same ${\rho_k^{(d)}(s,\Lambda)}$ obtained in \rro\ can be used
to reproduce
the momentum
cutoff structure, we consider again the simple scalar theory.
In this $\zeta$-function
cutoff formalism, the one-loop correction to $U_k$ is written as
\eqn\ukon{\eqalign{ U_k^{(1)}(\Phi) &={1\over 2}\int_p^{'}{\rm ln}
\Bigl({{p^2+V''(\Phi)}\over {p^2+V''(0)}}\Bigr) \cr
&
\longrightarrow -{1\over 2(4\pi)^{d/2}}\lim_{t\to 0}{d\over dt}\Biggl\{
{1\over\Gamma(t)}\int_0^{\infty}ds~s^{t-1-d/2}{\rho_k^{(d)}(s,\Lambda)}~\Bigl(
e^{-V''(\Phi)s}-e^{-V''(0)s}\Bigr)\Biggr\}.}}
By demanding that \ukon\ yields the same differential flow equation
for $U_k$ as that obtained from momentum cutoff regularization,
we are lead to
\eqn\eqwe{\eqalign{ &\lim_{t\to 0}~{d\over dt}\Biggl\{{1\over\Gamma(t)}
\int_0^{\infty}ds~s^{t-1-d/2}\Bigl({k{\partial\rho^{(d)}_k(s,\Lambda)\over
{\partial k}}}\Bigr)\Bigl(e^{-V''(\Phi)s}
-e^{-V''(0)s}\Bigr)\Biggr\} \cr
&
=-{2k^d\over\Gamma(d/2)}\int_0^{\infty}{ds\over s}e^{-k^2s}
\Bigl(e^{-V''(\Phi)s}-e^{-V''(0)s}\Bigr).}}
One can then verify by direct substitution that the above expression
is indeed satisfied by the smearing function given in \rro.
Thus the same ${\rho_k^{(d)}(s,\Lambda)}$ can be used to bring
the momentum cutoffs into $\zeta$-function
regularization although this may seem redundant since no divergence is
encountered in this prescription. Nevertheless,
by retaining the cutoff scales,
the flow pattern of the theory may be explored in a lucid manner.

As an explicit demonstration of $\zeta$ function cutoff regularization,
we compute the one-loop contribution of the blocked potential in $d=4$
and obtain
\eqn\ukonn{\eqalign{ U_k^{(1)}(\Phi) &
= -{1\over 32\pi^2}\lim_{t\to 0}~{d\over dt}\Biggl\{
{1\over\Gamma(t)}\int_0^{\infty}ds~s^{t-3}\Bigl[(1+k^2s)e^{-k^2s}
-(1+\Lambda^2s)e^{-\Lambda^2s}\Bigr] \cr
&\qquad\qquad\qquad
\times\Bigl(e^{-V''(\Phi)s}-e^{-V''(0)s}\Bigr)\Biggr\} \cr
&
=-{1\over 32\pi^2}\lim_{t\to 0}~{d\over dt}\Biggl\{{1\over{(t-1)(t-2)}}\Bigl[
\bigl(k^2+V''(\Phi)\bigr)^{2-t}-\bigl(\Lambda^2+V''(\Phi)\bigr)^{2-t}\Bigr]\cr
&\qquad\qquad\qquad\qquad
+{1\over{t-1}}\Bigl[k^2\bigl(k^2+V''(\Phi)\bigr)^{1-t}-\Lambda^2
\bigl(\Lambda^2+V''(\Phi)\bigr)^{1-t}\Bigr]+\cdots\Biggr\} \cr
&
={1\over 64\pi^2}\Biggl\{V''(\Phi)\bigl(\Lambda^2-k^2\bigr)
+\Lambda^4~{\rm ln}\Bigl({1+{V''(\Phi)\over\Lambda^2}}\Bigr)
-k^4~{\rm ln}\Bigl({1+{V''(\Phi)\over k^2}}\Bigr) \cr
&\qquad\qquad\quad
+{V''(\Phi)}^2~{\rm ln}\Bigl({k^2+V''(\Phi)\over{\Lambda^2
+V''(\Phi)}}\Bigr)\Biggr\}+\cdots.}}
Here we see that by keeping the UV cutoff $\Lambda$
finite when integrating over $s$,
counterterms have to be introduced to remove the
$\Lambda$-dependence.
On the other hand, if $\Lambda$ is first sent to infinity before
the $s$ integration, one arrives at
\eqn\ukot{\eqalign{ U_k^{(1)}(\Phi) &
= -{1\over 32\pi^2}\lim_{t\to 0}~{d\over dt}\Biggl\{
{1\over\Gamma(t)}\int_0^{\infty}ds~s^{t-3}(1+k^2s)e^{-k^2s}
\Bigl(e^{-V''(\Phi)s}-e^{-V''(0)s}\Bigr)\Biggr\} \cr
&
=-{1\over 32\pi^2}\lim_{t\to 0}~{d\over dt}\Biggl\{
{\bigl(k^2+V''(\Phi)\bigr)^{2-t}\over{(t-1)(t-2)}}
+{k^2\bigl(k^2+V''(\Phi)\bigr)^{1-t}\over {t-1}}+\cdots\Biggr\} \cr
&
=-{1\over 64\pi^2}\Biggl\{ k^2V''(\Phi)+{3\over 2}{V''(\Phi)}^2+\bigl(k^4
-{V''(\Phi)}^2\bigr){\rm ln}\Bigl({k^2+V''(\Phi)\over k^2}\Bigr)\Biggr\}
+\cdots,}}
which is precisely the finite one-loop contribution of the blocked potential.
It is interesting to note that taking the limit $\Lambda\to\infty$ before
and after the $s$ integration actually yields different results. In fact,
the two limits correspond to two different regularization procedures.
The connection between the $\zeta$-function cutoff
formalism and the momentum cutoff regularization actually be established
by the following integral transformation:
\eqn\ukey{\eqalign{ U_k^{(1)}(\Phi) &
= -{1\over 32\pi^2}\lim_{t\to 0}~{d\over dt}\Biggl\{
{1\over\Gamma(t)}\int_0^{\infty}ds~s^{t-3}{\rho_k^{(d)}(s,\Lambda)}~\Bigl(e^{-V''(\Phi)s}
-e^{-V''(0)s}\Bigr)\Biggr\} \cr
&
\longrightarrow -{1\over 2}\int_z^{'}
\lim_{t\to 0}~{d\over dt}\Biggl\{{1\over\Gamma(t)}\int_0^{\infty}ds~s^{t-1}
e^{-z^2s}\Bigl(e^{-V''(\Phi)s}-e^{-V''(0)s}\Bigr)\Biggr\} \cr
&
=-{1\over 2}\int_z^{'}
\lim_{t\to 0}~{d\over dt}\Biggl\{ {1\over \bigl(z^2+V''(\Phi)\bigr)^t}
-{1\over \bigl(z^2+V''(0)\bigr)^t}\Biggr\}
={1\over 2}\int_z^{'}~{\rm ln}\Bigl({{z^2+V''(\Phi)}
\over{z^2+V''(0)}}\Bigr).}}
In other words, equality with cutoff regularization is obtained by
keeping the $z$ integration till the end and interpreting $z$ as
the momentum scale $p$.

\goodbreak
\bigskip
\medskip
\goodbreak
\centerline{\bf REFERENCES}
\medskip
\nobreak
\medskip

\par\hang\noindent{\dimreg} G. 't Hooft and M. Veltman, {\it Nucl. Phys.}
{\bf B44} (1971) 189.
\medskip
\par\hang\noindent{\dowker} J. S. Dowker and R. Critchley, {\it Phys. Rev.}
{\bf D13} (1976) 3224.
\medskip
\par\hang\noindent{\pauli} W. Pauli and F. Villars, {\it Rev. Mod. Phys.}
{\bf 21} (1949) 434.
\medskip
\par\hang\noindent{\schwinger} J. Schwinger, {\it Phys. Rev.} {\bf 82}
(1951) 664.
\medskip
\par\hang\noindent{\michael} M. Oleszczuk, {\it Z. Phys.} {\bf C64}
(1994) 533.
\medskip
\par\hang\noindent{\ocr} S.-B. Liao, ``On the connection between momentum
cutoff and operator cutoff regularizations'', DUKE-TH-94-80/hep-th-9501124,
submitted to {\it Phys. Rev. D}.
\medskip
\par\hang\noindent{\chiral} R. D. Ball, {\it Phys. Rep.} {\bf 182} (1989) 1.
\medskip
\par\hang\noindent{\zuk} J. Zuk, {\it Int. J. Mod. Phys.} {\bf A18} (1990)
3549.
\medskip
\par\hang\noindent{\wilson} K. Wilson and J. Kogut, {\it Phys. Rep.}
{\bf 12C} (1975) 75; L. Kadanoff, {\it Physics} {\bf 2} (1966) 263.
\medskip
\par\hang\noindent{\blo} S.-B. Liao and J. Polonyi, {\it Ann. Phys.}
{\bf 222} (1993) 122 and {\it Phys. Rev.} {\bf D51} (1995) 4474.
\medskip
\par\hang\noindent{\ym} S.-B. Liao, {\it Chin. J. Phys.} {\bf 32} (1994) 1109;
and {\it Proceeding of the Workshop on Quantum Infrared Physics}, ed. by
H. M. Fried and B. M\"uller, (World Scientific, Singapore, 1995), p. 83.
\medskip
\par\hang\noindent{\wkrg} M. Bonini, M. DAttanasio and G. Marchesini,
{\it Nucl. Phys.} {\bf B437} (1995) 163 and {\bf B409} (1993) 441.
\medskip
\par\hang\noindent{\rt} M. Reuter and C. Wetterich,
{\it Nucl. Phys.} {\bf B417} (1994) 181 and {\bf 408} (1993) 91.
\medskip
\par\hang\noindent{\faddeev} L. D. Faddeev and A. A. Slavnov, {\it Gauge
fields} 2nd ed. (Benjamin, Reading, 1990).
\medskip
\par\hang\noindent{\nepomechie} R. I. Nepomechie, {\it Phys. Rev.} {\bf D31}
(1985) 3291; C. Mukku, {\it ibid.} {\bf D45} (1992) 2916.
\medskip
\par\hang\noindent{\duff} M. J. Duff and M. Ram\'on-Medrano,
{\it Phys. Rev.} {\bf D12} (1975) 3357.
\medskip
\par\hang\noindent{\abb} L. F. Abbot, {\it Nucl. Phys.} {\bf B185} (1981) 189;
A. Rebhan, {\it Phys. Rev.} {\bf D39} (1989) 3101.
\medskip
\par\hang\noindent{\mckeon} D. G. C. McKeon and T. N. Sherry,
{\it Ann. Phys.} {\bf 218} (1992) 325; {\it Phys. Rev. Lett.}
{\bf 59} (1987) 532; {\it Phys. Rev.} {\bf D35} (1987) 3854.
\medskip
\par\hang\noindent{\mike} S.-B. Liao and M. Strickland, {\it Phys. Rev.}
{\bf D52} (1995) 3653; S.-B. Liao, J. Polonyi and D. Xu,
{\it ibid.} {\bf D51} (1995) 748; S.-B. Liao
and J. Polonyi, {\it Nucl. Phys.} {\bf A570} (1994) 203c.
\medskip
\par\hang\noindent{\matinyan} S. G. Matinyan and G. K. Savvidy,
{\it Nucl. Phys.} {\bf B134} (1978) 539;
G. K. Savvidy, {\it Phys. Lett.} {\bf 71B} (1977) 133.
\medskip
\par\hang\noindent{\levi} A. R. Levi and J. Polonyi, {\it Phys. Lett.}
{\bf B357} (1995) 186; S. Huang and A. R. Levi, {\it Phys. Rev.}
{\bf D49} (1994) 6849.
\medskip
\par\hang\noindent{\maiani} see for example, L. maiani, G. Martinelli,
G. C. Rossi and M. Testa, {\it Nucl. Phys.} {\bf B175} (1980) 349.
\medskip
\par\hang\noindent{\nielsen} see for example, N. K. Nielsen,
{\it Am. J. Phys.} {\bf 49} (1981) 1171.
\medskip
\par\hang\noindent{\olsen}  N. K. Nielsen and P. Olsen, {\it Nucl. Phys.}
{\bf B144} (1978) 376;
J. Ambj\o rn and P. Olesen, {\it ibid.} {\bf B170} (1980) 60 and 265.
\medskip
\par\hang\noindent{\trottier} H. D. Trottier, {\it Phys. Rev.} {\bf D44}
(1991) 464; P. Cea and L. Cosmai, {\it ibid.} {\bf D48} (1993) 3364;
and H. D. Trottier and R. M. Woloshyn, {\it Phys. Rev. Lett.} {\bf 70}
(1993) 2053.
\medskip
\par\hang\noindent{\coleman} S. Coleman and E. Weinberg, {\it Phys. Rev.}
{\bf D7} (1973) 1888; M. R. Brown and M. J. Duff, {\it ibid.}
{\bf D11} (1975) 2124;
\medskip
\par\hang\noindent{\jackiw} R. Jackiw, {\it Phys. Rev.} {\bf D9} (1974) 1686.
\medskip
\par\hang\noindent{\sergei} S. G. Matinyan and G. K. Savvidy, {\it Sov. J.
Nucl. Phys.} {\bf 25} (1977) 118.
\medskip
\par\hang\noindent{\cheng} T.-P. Cheng and L.-F. Li, {\it Gauge theory
of elementary particle physics} (Clarendon, Oxford, 1984).
\medskip
\par\hang\noindent{\arfken} see for example, G. Arfken, {\it Mathematical
methods for physicists}, (Academic, New York, 1985).
\medskip
\par\hang\noindent{\ryder} see for example, L. Ryder, {\it Quantum field
theory}, (Cambridge, Cambridge, 1984), p. 397.
\medskip
\par\hang\noindent{\elizalde} E. Elizalde and J. Soto,
{\it Ann. Phys.} {\bf 162} (1985) 192; E. Elizalde and A. Romeo,
{\it Phys. Rev.} {\bf D40} (1989) 436 and {\it Int. J. Mod. Phys.}
{\bf A5} (1990) 1653.
\medskip
\par\hang\noindent{\gared} see for example, Z. X. Wang and D. R. Guo,
{\it Special functions}, (World Scientific, Singapore, 1989) p.107.
\medskip

\end